\documentclass[aps,12pt,axodraw,nofootinbib,superscriptaddress,aps]{revtex4}
\pdfoutput=1
\usepackage{epsfig}
\usepackage{amsmath}
\usepackage{bm}
\usepackage{times}
\usepackage{graphicx}
\usepackage{epstopdf}
\usepackage{amsfonts}
\usepackage{bm}
\usepackage{epsfig}
\usepackage{graphics}
\usepackage{xspace}
\usepackage[usenames]{color}

\def\bma#1{\mbox{\boldmath{$#1$}}}
\def\nn{\nonumber}
\def\bea{\begin{eqnarray}}
\def\eea{\end{eqnarray}}
\def\ba{\begin{eqnarray}}
\def\ea{\end{eqnarray}}
\def\be{\begin{equation}}
\def\ee{\end{equation}}

\def\beq{\begin{equation}}
\def\eeq{\end{equation}}

\def\nn{\nonumber}

\begin{document}

\title{\Large NSUSY fits}

\author{Jos\'e R. Espinosa}
\email{jose.ramon.espinosa.sedano@cern.ch}
\affiliation{ICREA at IFAE, Universitat Aut{\`o}noma de Barcelona, 08193 Bellaterra, Barcelona, Spain}

\author{Christophe Grojean}
\email{Christophe.Grojean@cern.ch}
\affiliation{ICREA at IFAE, Universitat Aut{\`o}noma de Barcelona, 08193 Bellaterra, Barcelona, Spain}
\affiliation{Theory Division, Physics Department, CERN, CH-1211 Geneva 23, Switzerland}

\author{Ver\'onica Sanz}
\email{vsanz@yorku.ca}
\affiliation{Theory Division, Physics Department, CERN, CH-1211 Geneva 23, Switzerland}
\affiliation{Department of Physics and Astronomy, York University, Toronto, ON, Canada, M3J 1P3}

\author{Michael Trott}
\email{michael.trott@cern.ch}
\affiliation{Theory Division, Physics Department, CERN, CH-1211 Geneva 23, Switzerland}

\date{\today}
\begin{abstract}
We perform a global fit to Higgs signal-strength data in the context of light stops in Natural SUSY.
 In this case, the Wilson coefficients of the higher dimensional operators mediating  $g \, g \rightarrow h$ and $ h \rightarrow \gamma \, \gamma$, given by $c_g,c_\gamma$, are related 
by $c_g = 3 \, (1 + 3 \, \alpha_s/(2 \, \pi)) c_\gamma/8$. We examine this predictive scenario in detail, combining Higgs signal-strength constraints  
with recent precision measurements of $m_W$, ${\rm Br}(\bar{B} \rightarrow X_s \, \gamma)$ constraints and direct collider bounds on weak scale SUSY, finding regions of parameter space that 
are consistent with all of these constraints. However it is challenging for the allowed parameter space to reproduce the observed Higgs mass value with sub-TeV stops. 
We discuss some of the direct stop discovery prospects and show how Higgs search data can be used to exclude light stop parameter space difficult to probe by direct collider searches. We determine the current status of such indirect exclusions and estimate their reach by the end of the $8 \, {\rm TeV}$ LHC run.
\end{abstract}
\maketitle
\section{Introduction}

The discovery of the Standard Model (SM) Higgs (assuming that the discovered boson is the SM Higgs) puts Supersymmetry (SUSY) to a new test. The possibility that SUSY solves the hierarchy problem without re-introducing fine-tuning i.e. the paradigm of Natural SUSY (NSUSY)~\cite{Dimopoulos:1995mi,Pomarol:1995xc,Cohen:1996vb,naturalsusy,
Kats:2011qh,Brust:2011tb,Papucci:2011wy} can now be examined in the light of increasingly precise measurements of  the properties of the Higgs. 

NSUSY predicts new particles near the electroweak (EW) scale, which necessarily must affect the properties of the Higgs if they are to stabilize this scale. 
The minimalistic scenario for NSUSY focuses on the vestige of the SUSY spectrum which is required to be light in order to keep the fine-tuning of the theory reasonably small. In this case,
the most significant impact of new states on Higgs phenomenology is through the presence of light stop states.
This scenario is not just motivated by simplicity, but also by the lack of evidence for SUSY to date, indicating that a weak scale SUSY spectrum needs to be non-generic to satisfy collider constraints. 
The states directly related to naturalness (primarily the stop and Higgsinos) are especially challenging, and model independent collider bounds are weak or non-existent.

Conversely, the study of the impact of an NSUSY scenario on the properties of the Higgs benefits from the enormous effort expended by the experimental
collaborations in refining the accuracy and precision of the reported Higgs signal strength measurements. When considering the experimentally resolvable impact of NSUSY,
indirect probes, {\it e.g.} through a fit to Higgs properties, electroweak precision data and flavor physics may well be more powerful in constraining many minimal scenarios than direct searches
for some time. This is the line of reasoning we develop in this paper, where we examine the current constraints on minimal NSUSY from these indirect probes.

The outline of this paper is as follows. We briefly review and introduce NSUSY in Section~\ref{NaturalSUSY}. In Section \ref{modifs} we then review the impact of the NSUSY spectra on the properties
of the SM Higgs through modifications in the loop-induced $h \rightarrow g \, g$ and $h \rightarrow \gamma \, \gamma$ couplings. Further, in Section~\ref{constraints} we work through the constraints set on  NSUSY from global fits to Higgs properties in the presence of light stops. We advance such studies  by using a more complete global fit  (now including 48 signal strength channels, including ICHEP and post ICHEP data updates)\footnote{Some past papers that have consistently examined earlier versions of the
Higgs dataset in this context, with varying degrees of sophistication, are Refs~\cite{Carena:2012gp,Carena:2011aa,Carmi:2012in,Carmi:2012yp, Cohen:2012zz,Curtin:2012aa,Arvanitaki:2011ck,Arbey:2011ab,jiji-new,craig-new,Boudjema:2012cq}.}. We make use of the fixed relationship between the Wilson coefficients (including the QCD matching corrections) for $h\gamma\gamma$ and $hgg$ in the case of NSUSY to perform a one-parameter fit and then directly map the allowed parameter space in global Higgs fits into the allowed stop space. Further, we determine $95 \%$ confidence level (C.L.) exclusion limits on the stop parameter space derived from Higgs search data. 
We then consider constraints from ${\rm BR}(\bar{B} \rightarrow X_s \, \gamma)$ and recent precision measurements of $m_W$ at the Tevatron,  as these results, which are under excellent theoretical control, are sensitive to light stops. The (statistically insignificant, but interesting) deviations from
the SM predictions in these observables in a weak scale NSUSY scenario could offer some further resolution on the allowed stop parameter space, if NSUSY exists. Ascribing these deviations to the effect of stops in NSUSY, 
such stop states are consistent with the results of the global fits to Higgs signal strengths, as we will show.
Finally, we also take into account direct collider bounds. In Section~\ref{combined} we discuss the interplay of these constraints and determine the allowed parameter space that remains. We include the limits coming from the Higgs mass measurement and estimate the degree of fine-tuning incurred, 
In Section~\ref{projections} we discuss the current exclusion bounds that can be derived using these indirect probes of stop parameter space. To study the future prospects of such limits by the end of 2012,  assuming that the experimental error in Higgs signal-strength measurements scales down as $\sim 1/\sqrt{\mathcal{L}_{int}}$, we consider two hypothetical cases: {\it 1)} the current pattern of best-fit signal-strength values does not change, and {\it 2)} the dataset evolves to converge on the SM expected signal strengths. Finally, in Section~\ref{concl} we conclude.

\section{Natural SUSY}\label{NaturalSUSY}
Naively, in generic SUSY scenarios motivated as a solution to the hierarchy problem, one expects all superpartners near the electroweak scale, with the soft breaking mass scale $\rm M_{SUSY}$ not higher than ${\cal O}(1)$ TeV.
\begin{table}[t] 
\setlength{\tabcolsep}{5pt}
\center
\begin{tabular}{c|c|c} 
\hline \hline 
Field & Spin & $\rm SU(3)_c \times SU(2)_L \times U(1)_Y$
\\
\hline
$\tilde{Q}_L = \left(\tilde{t}_L, \tilde{b}_L\right)$  & 0 & $({\bf 3}, {\bf 2}, 1/6)$
\\
$\tilde{t}_R^\ast$  & 0 & $({\bf \bar{3}}, {\bf 1}, -2/3)$
\\
$H_u  = \left(H_u^+ , H_u^0 \right)$  & 0 & $({\bf 1}, {\bf 2},+1/2)$ 
\\
$H_d  = \left(H_d^0 , H_d^- \right)$  & 0 & $({\bf 1}, {\bf 2},-1/2)$ 
\\
$\tilde{H}_{u}  = \left(\tilde{H}_{u}^+ , \tilde{H}_{u}^0 \right)$  & 1/2 & $({\bf 1}, {\bf 2},+1/2)$
\\
$\tilde{H}_{d}  = \left(\tilde{H}_{d}^0 , \tilde{H}_{d}^- \right)$  & 1/2 & $({\bf 1}, {\bf 2},-1/2)$
\\
$\tilde{g} $  & 1/2 & $({\bf 8}, {\bf 1},0)$ \\
\hline \hline
\end{tabular}
\caption{\it The minimal NSUSY field content. In our analysis, gluinos and a heavy linear combination of the Higgs doublets will be further integrated out.}
\label{table:SUSYtable} \vspace{-0.35cm}
\end{table}
The experimental picture emerging from the LHC is in growing tension with this expectation. After searching in many typical discovery channels, and reaching a peak sensitivity of roughly ${\cal O}( 1.5 {\rm TeV})$/ ${\cal O}(100)$'s GeV, for coloured/electroweak SUSY states~\cite{atlas-cms-general}, no statistically significant experimental excess has been reported to date. On the other hand, to avoid destabilizing the electroweak scale when  $\rm M_{SUSY} \gg {\it v}$ (without fine-tuning), only a minimal set of SUSY particles have to be light ($\lesssim 1-2  \, {\rm TeV}$) 
\cite{Dimopoulos:1995mi,Pomarol:1995xc,
Cohen:1996vb,naturalsusy,Kats:2011qh,Brust:2011tb,Papucci:2011wy}. 
The stop soft masses are directly connected (at one-loop) to the Higgs mass scale (or $Z$ mass) through the sizeable top coupling, so fine-tuning considerations require them to be light. Although
sbottoms ($\tilde{b}$) do not directly affect the fine-tuning of the Z mass, $\tilde{b}_L$ is required to be light as it is linked to the $\tilde{t}_L$ mass scale by $\rm SU(2)_L$ symmetry.  The Higgsino mass coming from the $\mu$ term\footnote{In this paper we distinguish
 signal-strengths with a subscript, $\mu_i$ for a final state $i$, from the $\mu$ parameter in NSUSY, which carries no subscript.} is tied to tree-level contributions to the Higgs mass. With gaugino masses assumed heavy, of order $M_{SUSY}$, light charginos and neutralinos are almost pure Higgsinos, with mass given by $\mu$ up to corrections that we neglect. 
To a lesser degree, the gluinos $\tilde{g}$ are also expected to be light due to their contribution to two-loop corrections to the Higgs mass parameter, which leads to the rough estimate $m_{\tilde{g}} \lesssim 2 m_{\tilde{t}}$ ($4 m_{\tilde{t}}$) for a Majorana (Dirac) gluino~\cite{Brust:2011tb}. 
In practice, we can also decouple these somewhat heavier gluinos in our analysis. 

The states just discussed are listed in Table \ref{table:SUSYtable}. These are the states whose impact on Higgs signal strengths and low-energy precision measurements we will focus on in this paper. We will not consider light staus ($\tilde{\tau}^{\pm}$), which can also modify the Higgs decays to photons, see Refs.~\cite{Carena:2012gp,jiji-new} for recent studies. We neglect these states as we assume a moderate value of $\tan\beta \lesssim {\cal O} (10)$, for which $\tilde{\tau}^{\pm}$ effects are negligible compared to the stop contribution. We also neglect the effects of the $\tilde{b}$ on the Higgs mass and in the loop corrections to the Higgs signal strength parameters for the same reason.

Regarding the SUSY Higgs sector, we will consider the decoupling regime in which only one Higgs doublet remains light\footnote{This choice is supported by the results of Ref.~\cite{craig-new}, where the 2HDM is studied in the light of the Post-Moriond Higgs data and no compelling region in the parameter space of the MSSM consistent with the data was identified, except the decoupling limit.}, while the second doublet, with mass controlled by the pseudoscalar mass $m_A$, has a mass $\sim {\rm TeV}$.
In this limit, the couplings of the lightest Higgs $h$ to fermions and gauge bosons approach their SM values, and we will only consider deviations in the loop-induced couplings of the light $h$ to photons and gluons (see next section).

In order to fix our notation, we write now the parts of the low-energy Lagrangian most relevant for our analysis. This Lagrangian is not supersymmetric as it applies below the scale of the heavy SUSY particles (with masses $\sim M_{SUSY}\gtrsim$ 1 TeV). Supersymmetric relations between some couplings are broken and one should introduce different couplings, to be matched to the supersymmetric theory at the scale $M_{SUSY}$. In practice,
the hierarchy between $M_{SUSY}$ and the electroweak scale is mild and one can neglect most of these breaking effects. 

From the superpotential $W=\mu H_d \cdot H_u + h_t 
Q_L \cdot H_2 t_R^c$, the Lagrangian gets the terms
\be
\label{Lagrangian1}
\delta {\cal L} = \mu \tilde{H}_d\cdot \tilde{H}_u +
h_t Q_L\cdot H_u t_R^c + h_t \tilde{Q}_L\cdot \tilde{H}_u 
t_R^c + h_t Q_L\cdot \tilde{H}_u \tilde{t}_R^* + {\rm h.c.} 
\ee
where $\cdot$ stands for the $\rm SU(2)_L$ product: $H_d\cdot H_u=H_d^0 H_u^0 - H_d^- H_u^+$ and we have suppressed
$\rm SU(3)$ indices for (s)quarks. (Weyl) fermionic fields are 
contracted in the usual way: $\tilde{H}_d^0\tilde{H}_u^0=i (\tilde{H}_d^0)^T\sigma_2\tilde{H}_u^0$.  Eq.~(\ref{Lagrangian1}) contains a Dirac mass term $\mu$ for Higgsinos,  the top Yukawa coupling between top quarks and
the Higgs, and the related Higgsino-squark-quark couplings, relevant for the contribution of Higgsino-stops to ${\rm Br}(\bar{B} \to X_s \gamma)$. 

The scalar potential for Higgses and squarks is well known and includes supersymmetric $F$ and $D$ terms and soft-SUSY-breaking terms. We can always perform a rotation of the full Higgs doublets $H_{u,d}$ to the doublets
$H_{l,h}$:
\be
\left(
\begin{array}{c}
H_l\\
\bar{H}_h
\end{array}
\right)=
\left(
\begin{array}{cc}
\cos\beta&\sin\beta\\
-\sin\beta & \cos\beta
\end{array}
\right)
\left(
\begin{array}{c}
\bar{H}_d\\
H_u
\end{array}
\right) \ ,
\ee
where $\tan\beta\equiv \langle H_u^0\rangle/\langle H_d^0\rangle= v_u/v_d$ and $\bar{H}_d$ denotes the $\rm SU(2)$ conjugate, 
$\bar{H}_d=-i\sigma^2 H_d^*=(-H_d^+,H_d^{0*})$, in such a way that $H_l$ is the doublet involved in electroweak symmetry breaking while $H_h$ does not take a vacuum expectation value. In the Higgs decoupling limit, with large pseudoscalar mass, $H_h$ is in fact composed of the heavy fields $H^0, A^0, H^\pm$, while
$H_l$ is SM-like and contains the light Higgs $h$ and the Goldstones. The quartic $H_l$ coupling determines the
light Higgs mass as usual and is the prime example of a coupling that receives sizeable SUSY-breaking corrections (that 
help in increasing the Higgs mass above its tree level minimal SUSY value below $m_Z$). Such corrections will be discussed in Subsection \ref{combined}.A.

Finally, the stop masses are given by the mass matrix
\be
{\cal M}^2_{\tilde{t}}=\left[
\begin{array}{cc}
M_{LL}^2&M_{LR}^2\\
M_{RL}^2&M_{RR}^2
\end{array}
\right]=
\left[
\begin{array}{cc}
M_{\tilde{Q}_L}^2+m_t^2+ m_Z^2 \left(\frac{1}{2}-\frac{2}{3}s_w^2\right)c_{2\beta} & m_t(A_t+\mu/\tan\beta)\\
m_t(A_t+\mu/\tan\beta) & M_{\tilde{t}_R}^2+m_t^2+ \frac{2}{3}m_Z^2s_w^2c_{2\beta} 
\end{array}
\right]\ ,\label{stopm}
\ee
where we used $c_{2\beta}=\cos 2\beta$, $m_t$ is the top mass and $M_{\tilde{Q}_L}, M_{\tilde{t}_R}, A_t$ are soft
SUSY-breaking masses.
The stop mixing angle $\theta_t$ relates the interaction eigenstates $\tilde{t}_{L,R}$ to the mass eigenstates $\tilde{t}_{1,2}$ by the rotation
\be
\left(
\begin{array}{c}
\tilde{t}_1\\
\tilde{t}_2
\end{array}
\right)=
\left(
\begin{array}{cc}
\cos\theta_{\tilde{t}}&-\sin\theta_{\tilde{t}}\\
\sin\theta_{\tilde{t}} & \cos\theta_{\tilde{t}}
\end{array}
\right)
\left(
\begin{array}{c}
\tilde{t}_L\\
\tilde{t}_R
\end{array}
\right) \ .
\label{rot}
\ee
The mixing angle $\theta_{\tilde{t}}$ is taken in the interval $(-\pi/2,\pi/2)$ 
and defined by
\be
\cos 2\theta_{\tilde{t}}\equiv\frac{M^2_{RR}-M^2_{LL}}{\sqrt{(M^2_{LL}-M^2_{RR})^2+4M_{LR}^4}}\ ,
\quad\quad
\sin 2\theta_{\tilde{t}}\equiv\frac{2M^2_{LR}}{\sqrt{(M^2_{LL}-M^2_{RR})^2+4M_{LR}^4}}
\ ,
\ee
with the signs of $M^2_{RR}-M^2_{LL}$ and $M^2_{LR}$
determining the quadrant of $2\theta_{\tilde{t}}$. With this definition
of $\theta_{\tilde{t}}$, one automatically guarantees
$m_{\tilde{t}_1}\leq m_{\tilde{t}_2}$, with
\be
m_{\tilde{t}_2}^2-m_{\tilde{t}_1}^2\equiv (\delta m)^2=\sqrt{(M^2_{LL}-M^2_{RR})^2+4M_{LR}^4}\ .
\ee
Finally, neglecting sbottom mixing (proportional to $m_b$), the light sbottom has mass $m_{\tilde{b}_L}^2=M_{\tilde{Q}_L}^2+m_b^2- m_Z^2(1/2-s_w^2/3)\cos2\beta$, while the heavy $\tilde{b}_R$ is decoupled.

\section{Loop-Level Corrections to Higgs properties in NSUSY} \label{modifs}

In this section, we review the loop-level NSUSY corrections to the couplings of $h$ to photons and gluons.\footnote{The effects of NSUSY loop corrections on the decay $\Gamma(h \rightarrow Z \, \gamma)$ are neglected as the leading effects in these scenarios come from charged scalars in the loop, and we are considering the decoupling limit in the scalar sector (consistent with the minimal version of NSUSY). Once $\mu_{Z\gamma}$ is reported further bounds on stops can be obtained through the related stop loop diagrams contributing to these decays.}
The leading correction from stops to the gluon-fusion process is given by \cite{Ellis:1975ap,Djouadi:1998az}
\bea \label{rgludef}
\frac{\sigma(gg \rightarrow h)}{\sigma^{SM}(gg \rightarrow h)} \simeq \frac{\Gamma(h \rightarrow gg)}{\Gamma^{SM}(h \rightarrow gg)} \simeq \left|1 +  r_g \right|^2,
\eea
where
\bea
r_g=\frac{C_g(\alpha_s) \, F_g(m_{\tilde{t}_1},m_{\tilde{t}_2},\theta_{\tilde{t}})}{F_g^{SM}(m_t,m_b \cdots)}  \ .
\eea
Here $C_g(\alpha_s)$ is a factor that takes into account higher order QCD corrections-- see the discussion below-- and the $F_g$ functions are defined as follows
\bea
F_g(m_{\tilde{t}_1},m_{\tilde{t}_2},\theta_{\tilde{t}}) &=&    \sum_{i=\tilde{t}_1,\tilde{t}_2 \cdots} g_{h \,i \, i}  \, \frac{m_Z^2}{m_{i}^2} \, F_0 (\tau_i), \\
F_g^{SM}(m_t,m_b \cdots) &=&  \sum_{i=t,b \cdots} F_{1/2}(\tau_i)\, \left(1 + \frac{11 \, \alpha_s}{4\, \pi}\right)
\approx -2/\left( 1.41 - 0.14  \, i \right),
\label{F}
\eea
where  $\tau_i = m_h^2/(4 m_{i}^2)$, $F_0(\tau)=\left[\tau -f(\tau) \right]/\tau^2$ and $F_{1/2}(\tau)= - 2 {\left[\tau + (\tau -1) f(\tau) \right]}/{\tau^2}$  with $ f(\tau)= \arcsin^2 \sqrt{\tau}$ for $\tau \leq 1$
while, for $\tau > 1$,
\bea
f(\tau)= -\frac{1}{4} \, \left[\log \frac{1 + \sqrt{1 - \tau^{-1}}}{1 - \sqrt{1 - \tau^{-1}}} - i \, \pi \right]^2.
\eea
See the Appendix for the SM inputs used in determining the numerical value of $F_g^{SM}$ above. 
The QCD correction applied above for the $b$ quark contribution is kept to the value of the correction quoted above, which is determined in the large quark mass limit $m_q \gg m_h$. The correction to this QCD correction due to the smaller $b$ quark mass is (very) subdominant in the numerical results.

The couplings $ g_{h \, \tilde{t}_i \, \tilde{t}_i}$ are given, in the decoupling limit by
\be
g_{h \tilde{t}_i \tilde{t}_i}\frac{m_Z^2}{m_t^2} =  1 + 
\langle \tilde{t}_i|\tilde{t}_L\rangle \frac{M_{LR}^2}{m_t^2} 
\langle \tilde{t}_R|\tilde{t}_i\rangle +
\frac{m_Z^2c_{2\beta}}{6m_t^2} \left[(3-4s_W^2)|\langle \tilde{t}_i|\tilde{t}_L\rangle|^2
+ 4s_W^2|\langle \tilde{t}_i|\tilde{t}_R\rangle|^2\right]
\ , 
\label{gtilde}
\ee
where $M_{LR}^2=m_t(A_t+\mu/\tan\beta)=(m_{\tilde{t}_2}^2-m_{\tilde{t}_1}^2)\sin\theta_{\tilde{t}}\cos\theta_{\tilde{t}}$ has been defined in Eq.~(\ref{stopm}); the $\langle \tilde{t}_i|\tilde{t}_{L,R}\rangle$ can be directly read from Eq.~(\ref{rot}); and we have included the $D$-term contributions proportional to  $\cos{2 \beta}$, although their effect is negligible.\footnote{Note that the stop contributions have an erroneous overall sign in Ref.~\cite{Djouadi:1998az} which propagated in the original version of this paper.}
The sign of $ \sum_{i=1,2} g_{h \tilde{t}_i  \tilde{t}_i} \,  \tau_i$ is positive if the stop sector is dominated by a light eigenstate of pure chirality, and negative if the term in $M_{LR}^2$ dominates, $M_{LR}^4\gtrsim 4m_t^2(m_{\tilde{t}_1}^2+m_{\tilde{t}_2}^2)$.
In the no-mixing case, we expect an enhancement of $\sigma(g g \to h )$. In the maximal-mixing case, the suppression depends on the separation between the two eigenstates. 

The decay width $\Gamma(h \rightarrow \gamma \, \gamma)$ is also modified by stop loops through the same function in Eq.~(\ref{F}), as the non-Abelian nature of QCD is irrelevant for the
leading-order loop function. One finds \cite{Ellis:1975ap} the correction
\bea
\frac{\Gamma(h \rightarrow \gamma \, \gamma)}{\Gamma^{SM}(h \rightarrow  \gamma \, \gamma)} \simeq \left|1 + r_{\gamma}  \right|^2, \quad \quad r_{\gamma} =\frac{N_c \, Q_{\tilde{t}}^2 \, C_\gamma(\alpha_s) F_g(m_{\tilde{t}_1},m_{\tilde{t}_2},\theta_{\tilde{t}})}{F_\gamma^{SM}(m_t,W,m_b \cdots) }.
\eea
The SM contribution is given by
\begin{equation}
F_\gamma^{SM}(m_t,W,m_b \cdots) = F_1(\tau_W)+\sum_{i=t,b \cdots} \! \!    N_c \, Q_i^2\,  F_{1/2}(\tau_i)  \left(1 - \frac{\alpha_s}{\pi} \right) \approx  1/(0.155 + 0.002  \, i)\ ,
\end{equation}
where $F_1 (\tau)= \left[ 2 \, \tau^2 + 3 \tau +  3( 2 \tau -1)  f(\tau) \right]/\tau^2$,
$N_c$ is the number of colours, and $Q_i$ is the electric charge with $e$ factored out. The matching correction in this case is 
given by $C_\gamma(\alpha_s)$, to be further discussed in the next section. 



%

\section{Current Constraints on NSUSY}\label{constraints}

\subsection{Effective Theory Approach}

It is useful to consider the approximation that all of the light NSUSY states are still heavy enough to be integrated out giving local operators. 
We can then fit to the data directly using the effective Lagrangian\footnote{We do not include
CP violating operators, assuming that all non-SM CP violating phases of the states integrated out are negligible. It has been argued \cite{Brust:2011tb} that this can be naturally accomplished in NSUSY when additional assumptions are employed concerning R-parity for example. Our assumption is conservative as the existence of large CP violating phases would only increase the constraints on a NSUSY spectrum.}
\bea
\mathcal{L}_{HD} &=& - \frac{c_g \, g_3^2}{2 \, \Lambda^2} \, H^\dagger \, H \, G^A_{\mu\, \nu} G^{A \, \mu \, \nu} - \frac{c_W \, g_2^2}{2 \, \Lambda^2} \, H^\dagger \, H \, W^a_{\mu\, \nu} W^{a \, \mu \, \nu} 
 - \frac{c_B \, g_1^2}{2 \, \Lambda^2} \, H^\dagger \, H \, B_{\mu\, \nu} B^{\mu \, \nu}, \nn \\
&\,&   - \frac{c_{WB} \, g_1 \, g_2}{2 \, \Lambda^2} \, H^\dagger \, \tau^a \, H \, B_{\mu\, \nu} W^{a \, \mu \, \nu}\ ,
\label{HDL}
\eea
where $g_1,g_2,g_3$ are the weak hypercharge, $\rm {SU}(2)$ gauge and $\rm{SU}(3)$ gauge couplings and the scale $\Lambda$ corresponds to the mass of the NSUSY states integrated out. 
The effects in NSUSY appear at the loop level, so we find it convenient to rescale the Wilson coefficients as $c_j = \tilde{c}_j/(16 \pi^2)$.
In this case, using the results of Ref. \cite{Manohar:2006gz}, the effect of the operators in (\ref{HDL}) is 
\bea\label{higher-d.effect}
\sigma_{gg \rightarrow h} \approx \sigma^{SM}_{gg \rightarrow h} \, \left| 1 +\frac{2}{F_{g}^{SM}} \frac{v^2 \, \tilde{c}_{g}}{\Lambda^2}\right|^2,   \quad \! \!
\Gamma_{h \rightarrow \gamma \, \gamma} \approx \Gamma^{SM}_{h \rightarrow \gamma \, \gamma} \, \left| 1 + \frac{1}{F_{\gamma}^{SM}}  \frac{v^2 \, \tilde{c}_{\gamma}}{\Lambda^2}\right|^2.  
\eea
Here $\tilde{c}_{\gamma} = \tilde{c}_W + \tilde{c}_B - \tilde{c}_{WB}$. 
%
%
%
We can translate the effect of stops in the language of local operators by inspecting our expressions in Sec.~\ref{modifs} and those in Eq.~(\ref{higher-d.effect}),
 \bea
 \frac{v^2 \, \tilde{c}_{g}}{\Lambda^2} \simeq C_g(\alpha_s) \, \frac{F_{g}}{2}, \quad \quad \, \frac{v^2 \, \tilde{c}_{\gamma}}{\Lambda^2} \simeq  N_c \, Q_{\tilde{t}}^2 \, C_\gamma(\alpha_s) \, F_{g}  \ .
 \eea
These relationships are only approximate in the sense that the limit $m_{\tilde{t}_{1,2}} \gg m_h$ should be taken in the loop functions $r_{g},r_{\gamma}$ to
match onto the local operators. If the two stop mass eigenstates can be integrated out simultaneously, the matching can be directly performed by expanding the loop functions in this limit,
and one obtains\footnote{Here we have neglected D term contributions, although we retain the effect of D terms in some of the numerical results presented. These corrections are negligible except for
$m_{\tilde{t}_1} < 150 \, {\rm GeV}$ masses. They introduce a (minor) $\tan \beta$ dependence into the definition of $F_g$ in the local operator approximation when retained.}
\bea
F_g = -\frac{1}{3} \, \left[\frac{m_t^2}{m_{\tilde{t}_1}^2} + \frac{m_t^2}{m_{\tilde{t}_2}^2}  - \frac{1}{4} \, \sin^2 (2 \, \theta_t) \, \frac{\delta m^4}{m_{\tilde{t}_1}^2 \, m_{\tilde{t}_2}^2}\right].
\eea
While, in this limit, the QCD matching corrections take the simple form 
\bea
C_g(\alpha_s) = 1 + \frac{25 \, \alpha_s}{6 \, \pi}, \quad \quad \quad C_\gamma(\alpha_s) = 1 + \frac{8 \, \alpha_s}{3 \, \pi}.
\eea
These perturbative corrections are the matching corrections due to top squarks in the loops that do not cancel when a ratio is taken with the SM contribution to these loops.
This correction factor is obtained in Ref.~\cite{Dawson:1996xz} in the limit where
gluino effects and the effect of squark mixing was neglected.\footnote{It has been pointed out that  $m_{\tilde{g}} \rightarrow \infty$ leads to mixed stop-gluino UV divergences \cite{Harlander:2003bb} requiring extra counter-terms, but this technical requirement is not a barrier to the
numerical investigations we perform.
The full matching correction is given in Ref.~\cite{Harlander:2004tp}: the gluino contributions and stop mixing effects are a small correction to the $\sim 5 \%$ matching correction we consider.}
This approach also neglects running  that would sum large logs if a two stage matching was employed, integrating out each stop 
eigenstate in sequence.\footnote{There are also perturbative corrections to the matrix element of the local effective operator 
$h \, G^A_{\mu\, \nu} G^{A \, \mu \, \nu} $. These are common multiplicative factors, as are soft gluon re-summation effects, and
cancel in the ratios taken.} 

Whether one integrates out the stops and matches onto the local operator approximation or not, there is a relationship between the NP effects on $\sigma_{gg \rightarrow h}$ and $\sigma_{h \rightarrow \gamma \gamma}$ that is independent of the stop mass parameters in the minimal NSUSY limit.  In the local operator approximation, the relationship is simply
\bea
\frac{\tilde{c}_{g}}{\tilde{c}_{\gamma}} =\frac{1}{2N_c Q_{\tilde{t}}^2}\frac{C_g(\alpha_s)}{C_\gamma(\alpha_s)}=
 \frac{3}{8} \, \left(1 + \frac{3 \, \alpha_s}{2\, \pi}\right) ,
\label{rel-stop}
\eea
where we see how the ratio $\sim 3/8$ is determined by the stop quantum numbers. This is a consequence of assuming that the only BSM contribution  to both the $\gamma\gamma$ and $gg$ loops comes from stops. This strong relationship will be relaxed in less minimal 
scenarios. For example, light $\chi^\pm$'s with mass $m_{\chi_1}^2 \sim \mu^2$ (in the decoupling limit)  would in principle also contribute to the $\gamma \, \gamma$ loops. However, the Higgs couples to the higgsino as $h \, \tilde{W}^{\pm} \, \tilde{H}^{\mp}$ and a large mixing between wino and higgsino eigenstates would be required. As we are considering the large gaugino mass limit in NSUSY, $M_2 \gg \mu, v$,  this mixing scales as  $\sim  m_W^2  \,  \sin^2 \beta /(M_2^2)$ and is suppressed, so that we can neglect the chargino contribution to $h\gamma\gamma$.

We also utilize this effective Lagrangian to examine the issue of efficiency corrections to the $\mu_i$
when high-dimension operators are present. We find that such efficiency corrections to event rates are very small and neglect them. See the Appendix for details. 

\subsection{Global Fit To Higgs Signal Strengths}\label{globalfit}
In this section we describe our method and results for globally fitting to Higgs signal strength data
in the scenario discussed above. Here we only briefly review the fit procedure, the details of our fit method are given in Refs.~\cite{Espinosa:2012ir,Espinosa:2012vu,Espinosa:2012im}.\footnote{For other model-independent approaches to the determination of the Higgs couplings, see \cite{Carmi:2012yp,Azatov:2012bz,Montull:2012ik,Giardino:2012ww,Ellis:2012rx,Lafaye:2009vr,Englert:2011aa,Klute:2012pu,
concha,Azatov:2012ga,Carmi:2012in}.}
Our fit incorporates the recently released $7$ and $8$ TeV LHC data \cite{Wedtalk,talkfriday,atlasupdate,Chatrchyan:2012tx}, and the recently reported Tevatron Higgs results \cite{tevatronupdate}. The data we use is listed in the Appendix.
We fit to the available Higgs signal-strength data, 
\bea
\mu_i = \frac{[ \sum_j \, \epsilon_{ij} \, \sigma_{j \rightarrow h} \times {\rm Br}(h \rightarrow i)]_{observed}}{[ \sum_j \, \epsilon_{ij} \,  \sigma_{j \rightarrow h} \times {\rm Br}(h \rightarrow i)]_{SM}}\ ,
\label{mui}
\eea 
for the production of a Higgs that decays into the observed channels $i = 1 \cdots N_{ch}$. Here $N_{ch}$ denotes the number of channels, 
the label $j$ in the cross section, $\sigma_{j \rightarrow h}$,  is due to the fact that some final states are summed over different Higgs production processes, labelled with $j$. The
$\epsilon_{ij}$ are the efficiency factors for the various production processes producing a final state $j$ to pass experimental cuts.
The reported best-fit value of a signal strength we denote by $\hat{\mu}_i$, and the $\chi^2$ we construct is defined as
\bea
\label{chi2}
\chi^2(\mu_i) =\sum_{i=1}^{N_{ch}}\frac{(\mu_i-\hat\mu_i)^2}{\sigma_i^2}\;.
\eea 
The covariance matrix has been taken to be diagonal with the square of the $1 \, \sigma$ theory and experimental errors added in quadrature giving $\sigma_i$.  We necessarily neglect correlation
coefficients as these are not supplied. For the experimental errors we use $\pm$ symmetric $1 \sigma$ errors on the reported $\hat{\mu}_i$, while  for theory predictions and related errors
we use the results of the LHC Higgs Cross Section Working Group \cite{Dittmaier:2011ti}.
The minimum ($\chi^2_{min}$) is determined, and the $68.2 \% \, (1 \, \sigma), 95 \% \, (2 \, \sigma), 99 \% \, (3 \, \sigma)$ best fit regions are plotted as $\chi^2 = \chi^2_{min} + \Delta \chi^2$, with
the appropriate cumulative distribution function (CDF) defining the corresponding $ \Delta \chi^2$.

\begin{figure}[h!]
\centering
\includegraphics[scale=0.8]{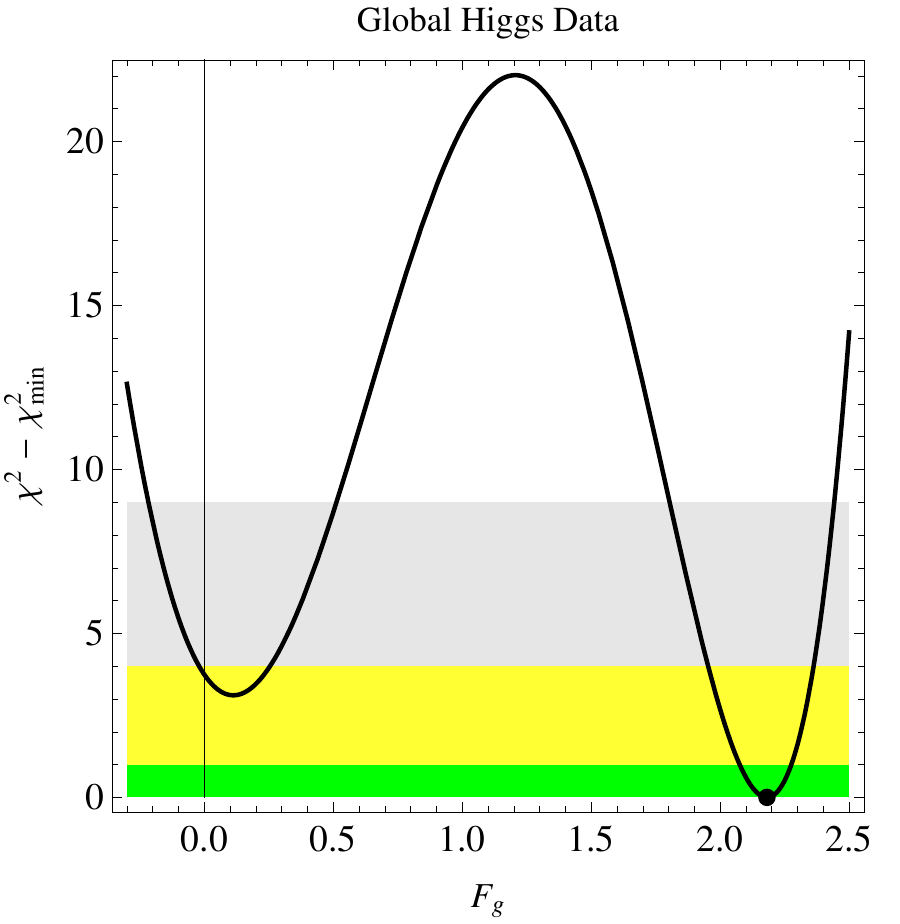}
\includegraphics[scale=0.8]{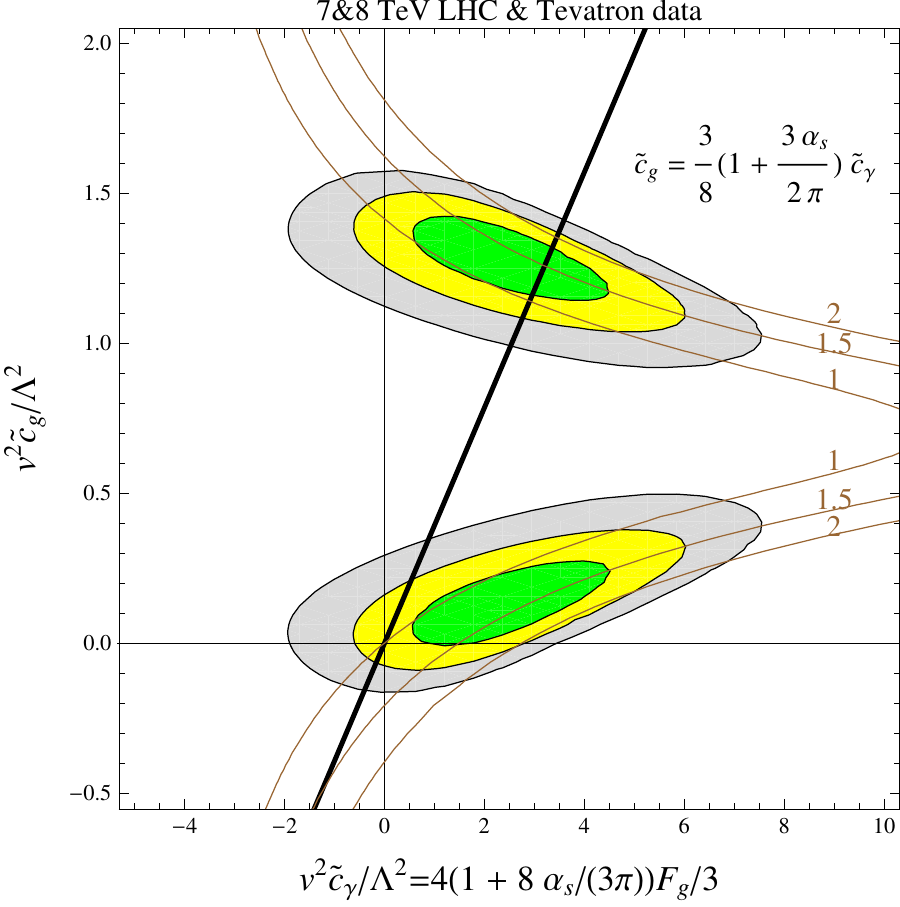}
\caption{The one dimensional fit to $F_g$ in the NSUSY scenario to global Higgs data (left), and the approximate projection of the relationship between
the Wilson coefficients into the higher dimensional operator space (right).The green, yellow and gray regions correspond to the $1,2,3 \, \sigma$ allowed regions in the 1D or 2D fit space (defined with the  CDF appropriate to each case.  This difference accounts for the mismatch in the $\Delta \chi^2$'s that define
the best-fit regions). Also shown as solid (brown) contours is the enhancement of the $\mu_{\gamma \, \gamma}$ signal strength and how such a condition projects into
the best fit space.}
\label{one-param}
\end{figure}

We first perform a one-parameter fit in terms of the free parameter $F_g$, that depends on the stop sector of NSUSY,  and plot $\chi^2-\chi^2_{min}$ in Fig.~\ref{one-param} (left). This $\chi^2$ distribution 
is directly related to the fit in the broader space of  the generic Wilson coefficients of the local operators contributing to $h\to\gamma\gamma$ and $gg\to h$, as shown in Fig.~\ref{one-param} (right),
although the match of the C.L. regions is only approximate due to the difference in number of degrees of freedom in the fit.
It is not surprising that with the addition of this free parameter, the $\chi^2$ measure is improved  as compared to the SM. However, it was by no means guaranteed that the stop line determined by the
NSUSY relationship between the Wilson coefficients would pierce the best fit region away from the
SM one. This accidental fact allows a $\sim 2 \sigma$ improvement of the fit. Although this is intriguing, 
we caution the reader that the interval of Wilson coefficients that intersect the $1 \, \sigma$ best fit region
is mapped into a very narrow range of stop mass parameters, corresponding to a tuned area of parameter space. In addition, best-fit regions in $(\tilde{c}_\gamma,\tilde{c}_g)$ space more distant from the SM point at $(0,0)$ will generically correspond to lighter states, as their impact scales as $1/\Lambda^2$. This will represent a further problem for this region.

We can characterize the allowed relationship between the Wilson coefficients that intersect the ($1 \sigma$) best fit region in a model-independent way, finding that current data is consistent with the following four ranges of the Wilson coefficient ratios, corresponding to the four different best-fit regions in that
2D space~\cite{Espinosa:2012im}.
\footnote{Note that these bounds are approximate in the following sense: for a 1D fixed relationship between the Wilson coefficients, the allowed C.L. regions are slightly different if obtained 
with the 1D CDF or for the 2D  Wilson coefficient case. Again, this effect can be seen in the NSUSY case in Fig.~\ref{one-param}} For $\tilde{c}_\gamma>0$:
\be
-0.01 \,  <  \tilde{c}_g /\tilde{c}_\gamma< 0.16 \, ,  \quad  \quad \quad \quad \quad 0.27 \, <  \tilde{c}_g/\tilde{c}_\gamma  < 2.5 \, , 
\ee
and, for $\tilde{c}_\gamma<0$:
\be
-0.1 \,  <  \tilde{c}_g / \tilde{c}_\gamma< - 0.065 \, ,  \quad  \quad \quad -0.016  <  \tilde{c}_g/\tilde{c}_\gamma < 0.001\ .
\ee

\begin{figure}
\centering
\includegraphics[scale=0.42]{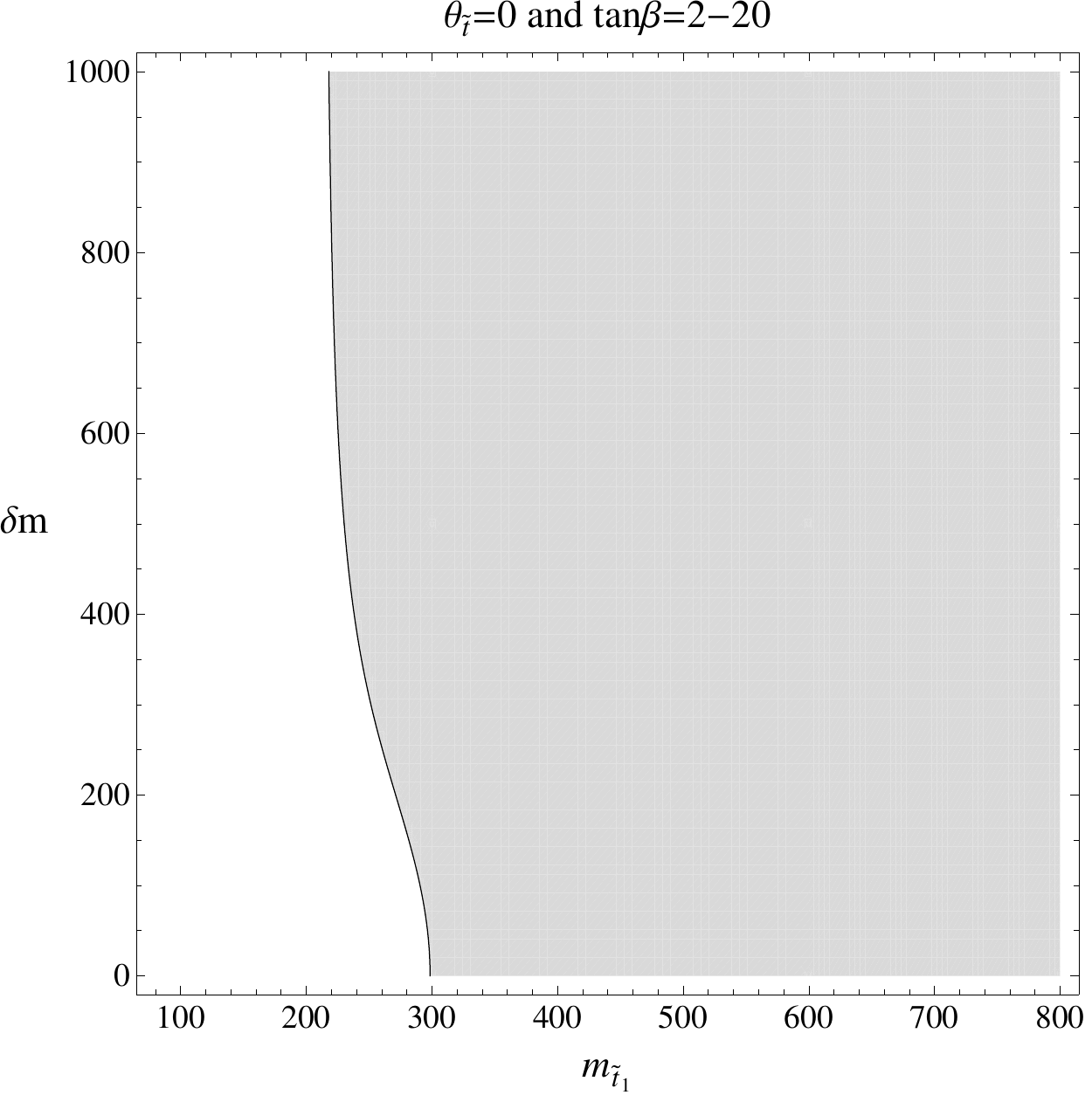}
\includegraphics[scale=0.42]{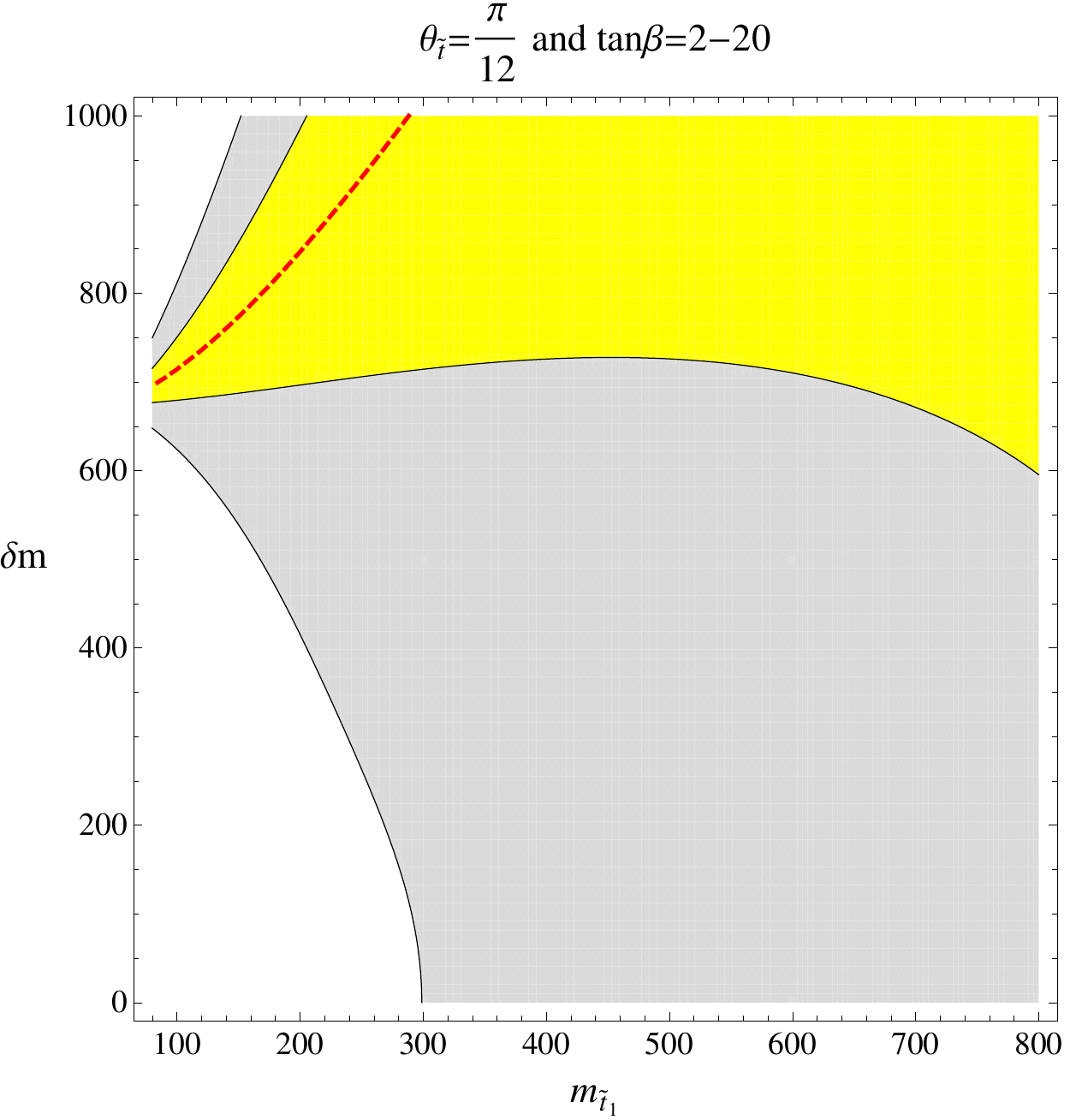}
\includegraphics[scale=0.42]{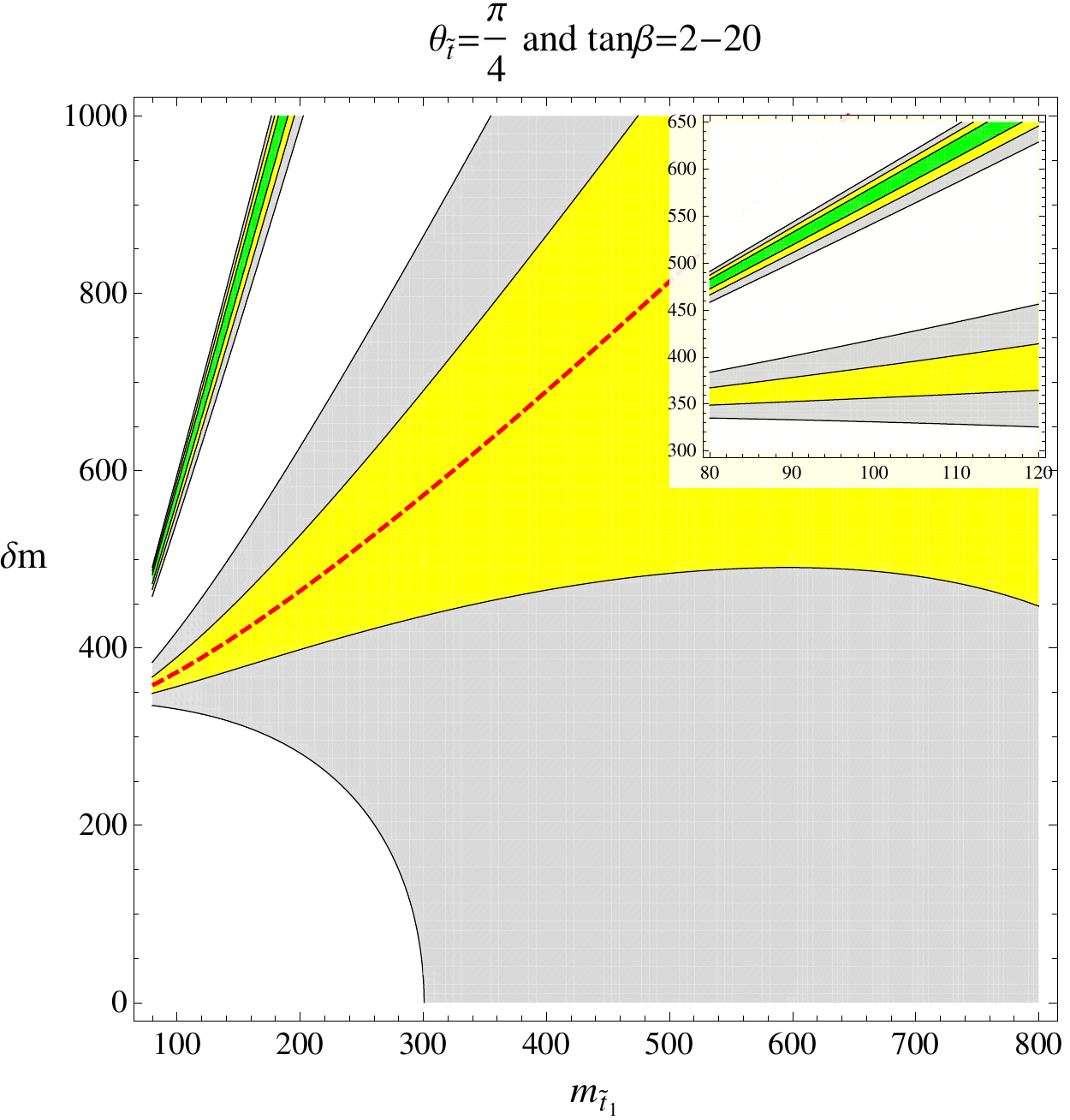}
\caption{The projection into stop parameter space of the best-fit regions from a global fit to Higgs signal strength data. Here $\delta m = (m_{\tilde{t}_2}^2 - m_{\tilde{t}_1}^2)^{1/2}$.
Colour convention is the same as in previous plots. The inset zooms into the low mass best-fit region 
(ruled out by LHC monophoton searches). We have varied $ \tan \beta$ in the range $(2,20)$ and taken the overlap of the best fit spaces, slightly increasing the allowed parameter space. 
The three plots show the cases of no mixing ($\theta_{\tilde{t}} = 0$), intermediate mixing ($\theta_{\tilde{t}} = \pi/12$) and maximal mixing ($\theta_{\tilde{t}} = \pi/4$). 
The dashed line corresponds to the second minimum in the one parameter $\chi^2$ shown in the previous figure.}
\label{stopspace}
\end{figure}

In the limit of a single field contributing to the Wilson coefficients, the $\tilde{c}_g/\tilde{c}_{\gamma}$ ratio is dictated by the quantum numbers of the field integrated out.\footnote{See Ref.~\cite{Bonnet:2012nm} for a recent study that also emphasizes this point.} Clearly, the study of the possible intersections of such lines with the best-fit  regions in the space $(\tilde{c}_g,\tilde{c}_{\gamma})$ for any model (including NSUSY)  will become much more important with further refinements in the measurement of Higgs properties. We discuss some prospects for the improvement of these fits in Section \ref{projections}.

For the NSUSY case, 
the light stop best-fit region occurs for $F_g \sim 2$; one can see how this space relates to the $(\tilde{c}_g,\tilde{c}_{\gamma})$ plane in Fig.~\ref{one-param}, and it corresponds to having the lightest stop mass eigenstate significantly lighter than the second stop eigenstate.
Most of this space is already strongly constrained by monophoton searches, as we discuss further below. NSUSY hopes in light of current global Higgs data (when our assumptions are adopted) are based on the consistency of NSUSY
in the $(\tilde{c}_g,\tilde{c}_\gamma)$ parameter space near the SM point $(\tilde{c}_g,\tilde{c}_{\gamma}) = (0,0)$, for larger $m_{\tilde{t}_1}$ and small $F_g$. Translating the allowed fit space to the space of the stop parameters is very convenient to discuss the interplay with further constraints and direct stop discovery prospects. When we translate the results of the global fit to Higgs signal-strengths to the stop space, we find the best-fit regions shown in Fig.~\ref{stopspace}. 
The three plots show the cases of no mixing ($\theta_{\tilde{t}} = 0$), intermediate mixing ($\theta_{\tilde{t}} = \pi/12$) and maximal mixing ($\theta_{\tilde{t}} = \pi/4$). We will show these canonical parameter choices throughout this paper when examining the global constraint picture.
The point in the 1D fit that has a local minimum $\chi^2$  with small $F_g$ is mapped into
the red dashed line in the two rightmost plots in  Fig.~\ref{stopspace}. It can be checked that the isocontours of fixed value of $F_g$ all converge to $m_{{\tilde t}_1}=0, \delta m^2=4m_t^2/\sin^2(2 \theta_{\tilde t})$, which explains that, for small values of the mixing angle, the red dashed line corresponds to a large spliting of the stops and for zero mixing the red dashed line does not exist since $F_g$ is always negative. As anticipated, the best-fit region at low
stop-mass is extremely narrow, and even when a disconnected region in the stop space exists, the region is surely fine-tuned.

\subsection{$\bma{ {\rm Br}(\bar{B} \rightarrow X_s \, \gamma)}$}

Another important challenge for the parameter space of NSUSY scenarios comes from non-SM contributions to magnetic moment operators. Although the contributions to these operators
vanish \cite{Ferrara:1974wb} in the pure SUSY limit,  NSUSY scenarios are far from this limit by construction. 
As a result, the reduction in the allowed parameter space due to constraints from ${\rm Br}(\bar{B} \rightarrow X_s \, \gamma)$ can be significant. 
Recall that the effective Lagrangian (neglecting light quark masses) is given by \cite{Grinstein:1987vj}
\bea
\mathcal{L}_{eff} &=&  \frac{4 \, G_F}{\sqrt{2}} \, V_{tb} \, V_{ts}^\star \, \sum_{i=1}^8 \, C_i(\mu) \, Q_i, \nn \\
&=& \frac{G_f}{4 \, \sqrt{2} \, \pi^2} \, V_{tb} \, V_{ts}^\star \, m_b \, \left[  C_7 \, \bar{s}_L \, \sigma^{\mu \, \nu} \, b_R \,  e \, F_{\mu \, \nu} + C_8 \, \bar{s}_L \, \sigma^{\mu \, \nu} \, T_a \, b_R \, g_s \, G^a_{\mu \, \nu} + \cdots \right] .
\label{bsgops}
\eea

The SUSY contributions to the magnetic moment operators are well known \cite{Bertolini:1990if}
and can be applied to the particular NSUSY scenario. The dominant contributions to the Wilson coefficients come from stop-chargino loops:
\bea
\Delta C_{7,8} \simeq \sum_{i =1}^2 \left\{ -|\langle \tilde{t}_i|\tilde{t}_R\rangle|^2 \frac{m_t^2}{3 \, s_{\beta}^2\, m_{\tilde{t}_i}^2} \, F^1_{7,8}\left[\frac{m_{\tilde{t}_i}^2}{m_{\chi_1}^2}\right] - 
\langle \tilde{t}_i|\tilde{t}_L\rangle\langle \tilde{t}_R|\tilde{t}_i\rangle \, \frac{m_t}{s_{2 \beta}\, m_{\chi_1}} \, F^3_{7,8}\left[\frac{m_{\tilde{t}_i}^2}{m_{\chi_1}^2}\right] \right\}\ . 
\eea
We have only retained the light $\chi_1^\pm$ with mass $m_{\chi_1^\pm} \simeq \mu$, which is consistent with NSUSY assumptions  (the gaugino mass is  $M_2  > \mu \gtrsim  m_W$) . 
The loop functions $F_{7,8}^j$ are given in the Appendix. We vary the $\mu$ parameter in the range $\sim 100 - 200 \, {\rm GeV}$. The lower limit of this range is set by LEP
bounds on $\tilde{\chi}^\pm$ \cite{pdg}; the upper limit by naturalness considerations \cite{Brust:2011tb}.
In principle, there are also loop contributions from the light $\chi^0$. However, these contributions can be strongly suppressed
in the minimal NSUSY limit we consider. For example, there is no source of breaking of the residual $\rm U(1)_R$ symmetry, 
unless Majorana masses for the gluino are introduced. This symmetry plays a role in suppressing effects of large $\tan\beta$ or 
proton decay or flavour violating observables which require a flip in chirality, see Refs.~\cite{muless,MRSSM}. 
We will consider a minimal flavour violating scenario  \cite{Chivukula:1987py,Hall:1990ac,D'Ambrosio:2002ex,Buras:2003jf,Cirigliano:2005ck} when examining the NSUSY spectrum, this can follow from the argument above directly. In the
case of Majorana gluino masses we assume MFV.
\begin{figure}
\centering
\includegraphics[scale=0.65]{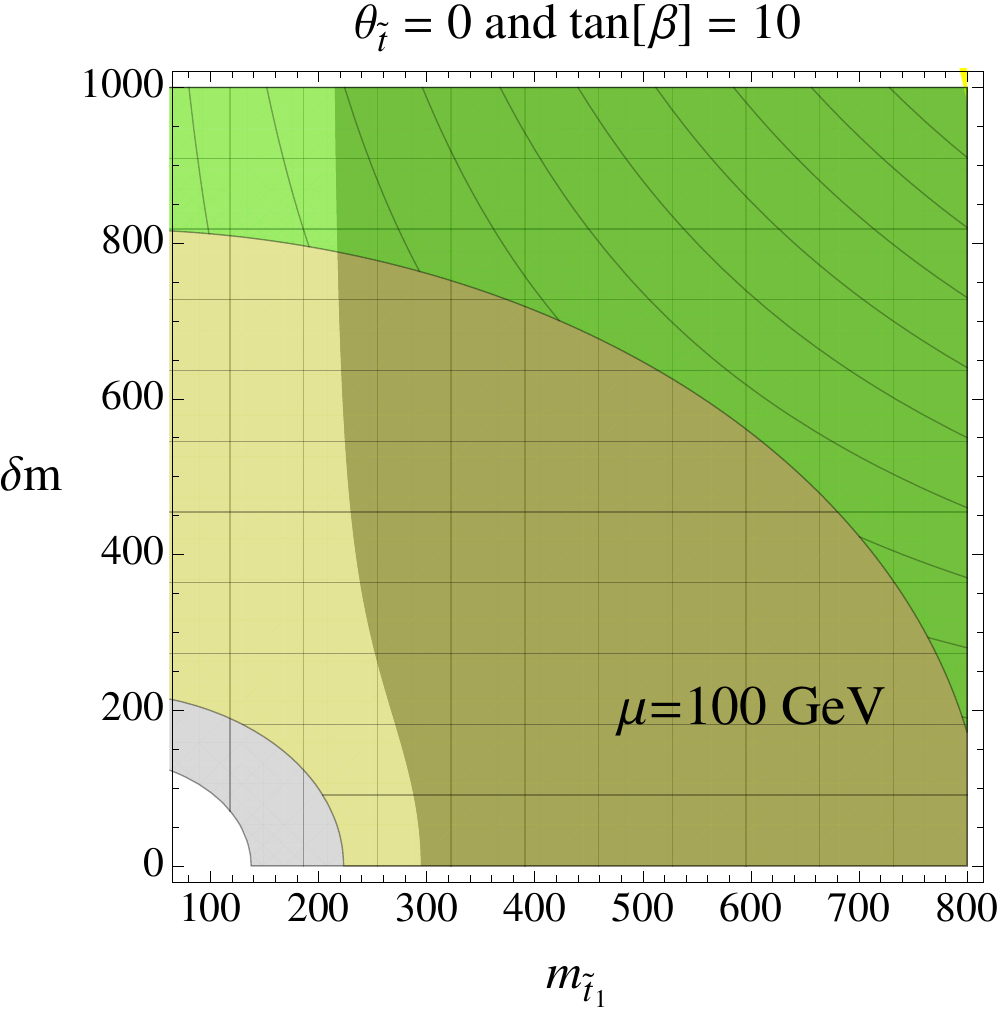}
\includegraphics[scale=0.65]{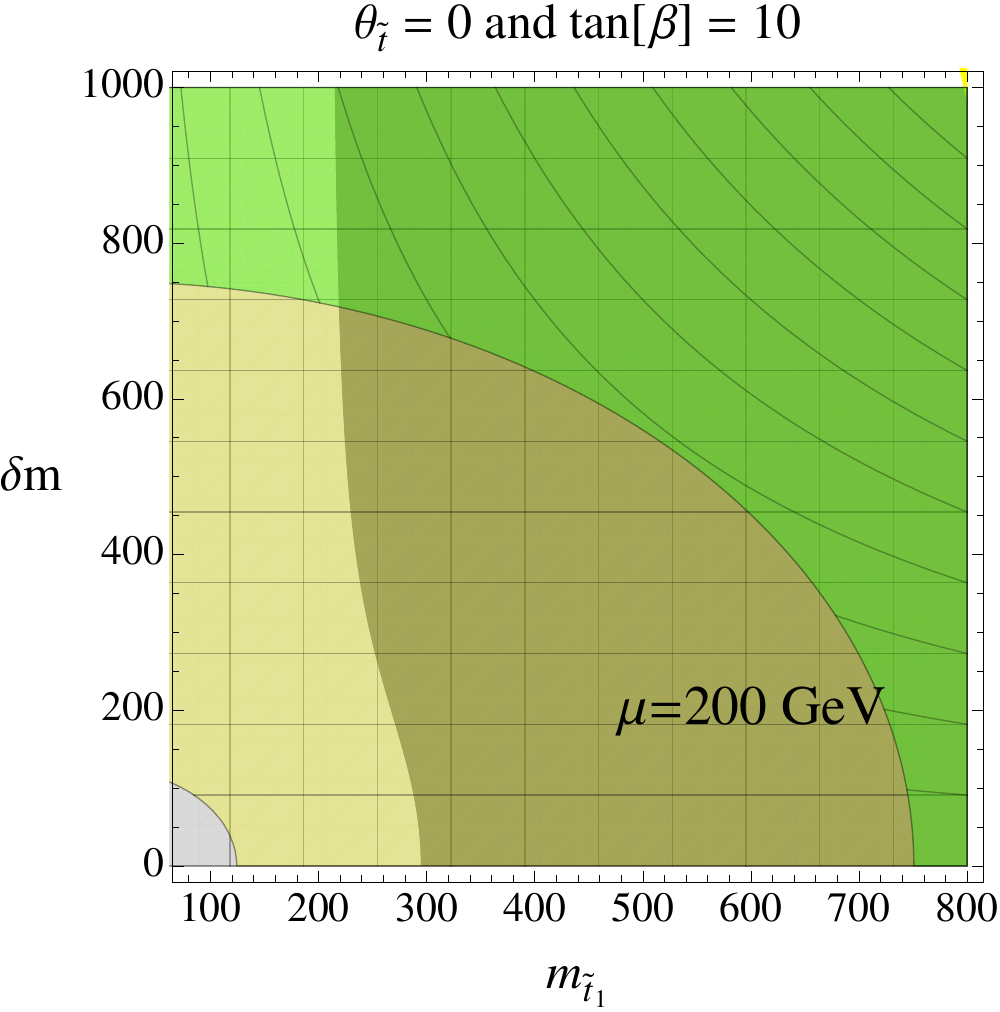}
\includegraphics[scale=0.65]{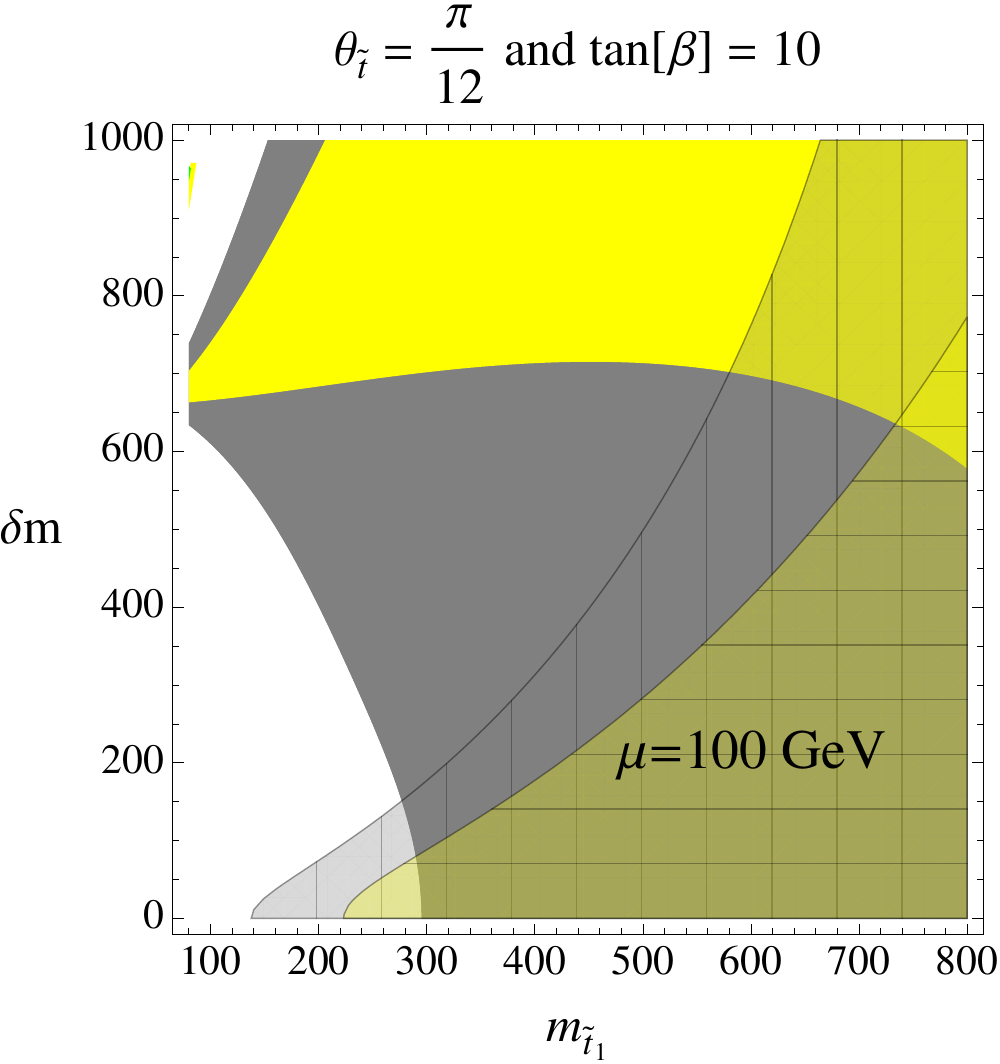}
\includegraphics[scale=0.65]{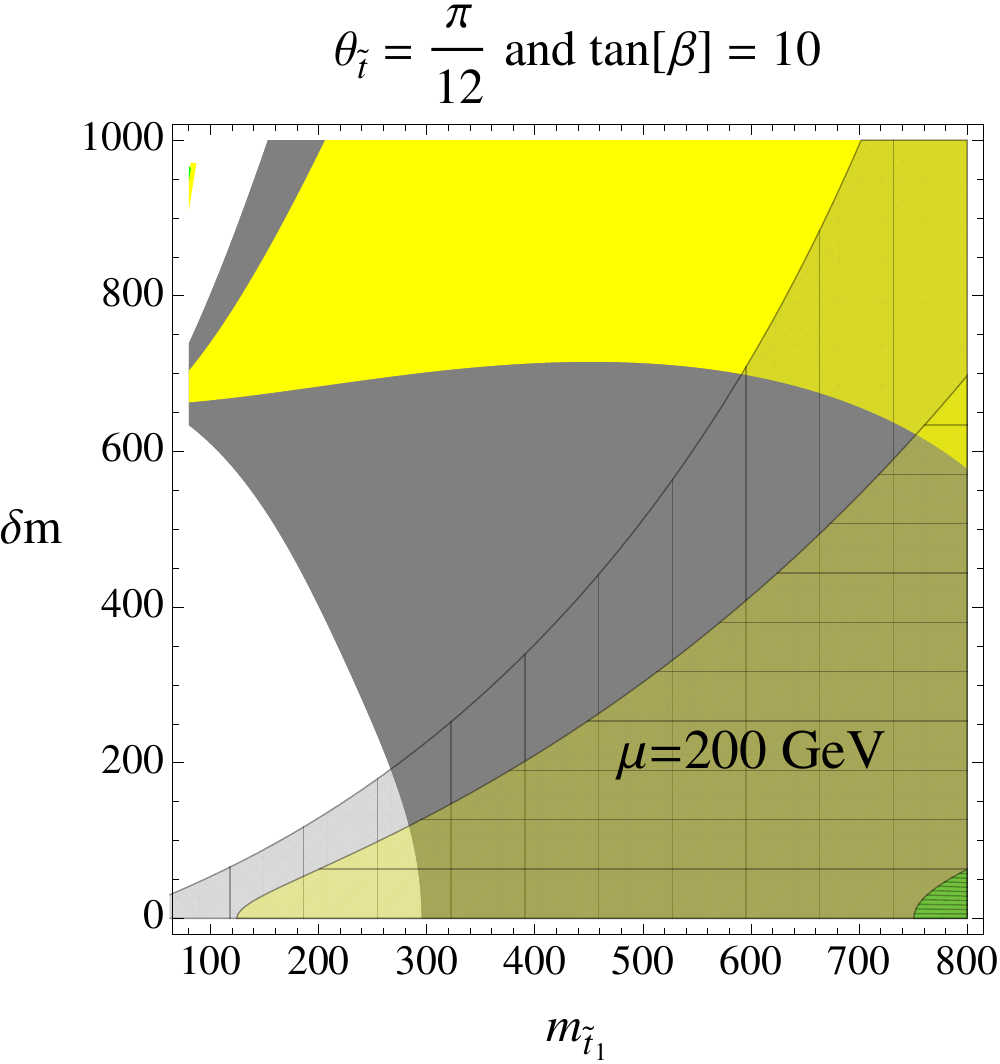}
\includegraphics[scale=0.65]{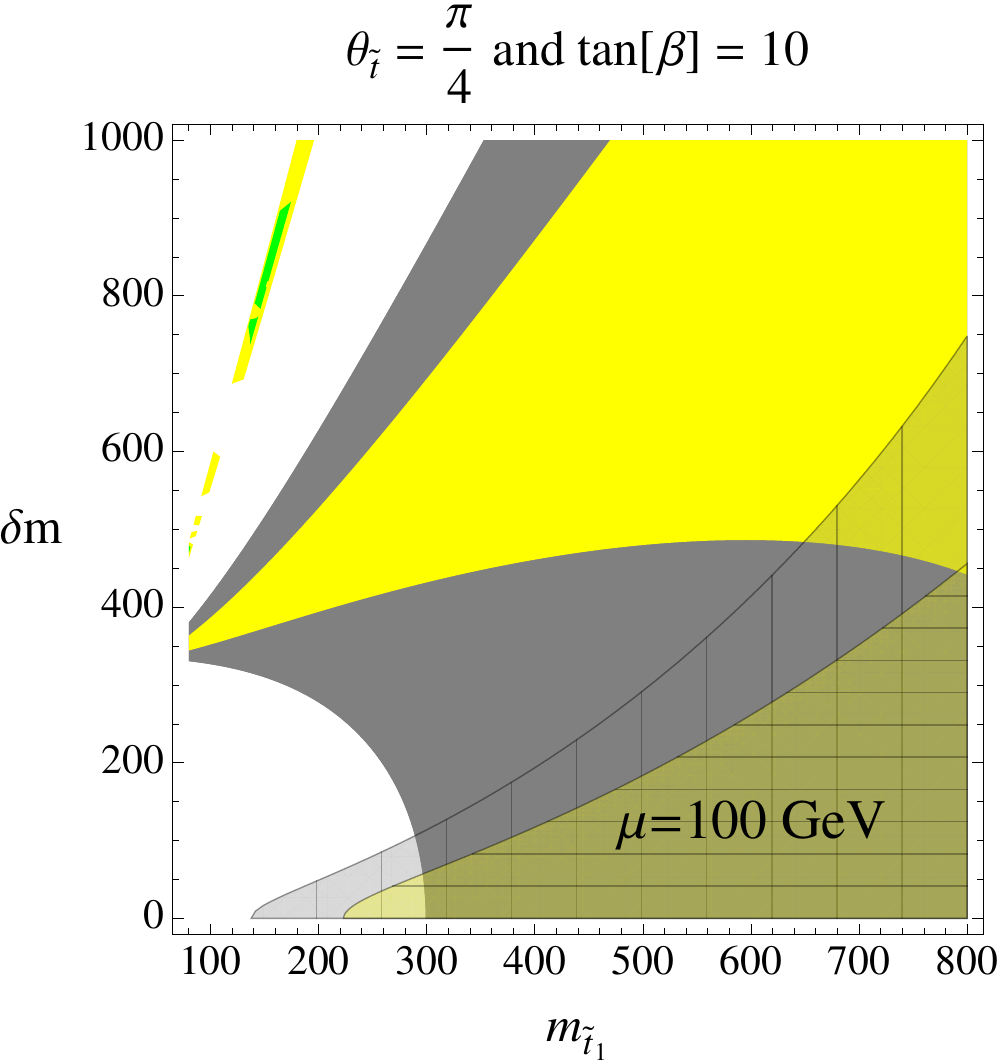}
\includegraphics[scale=0.65]{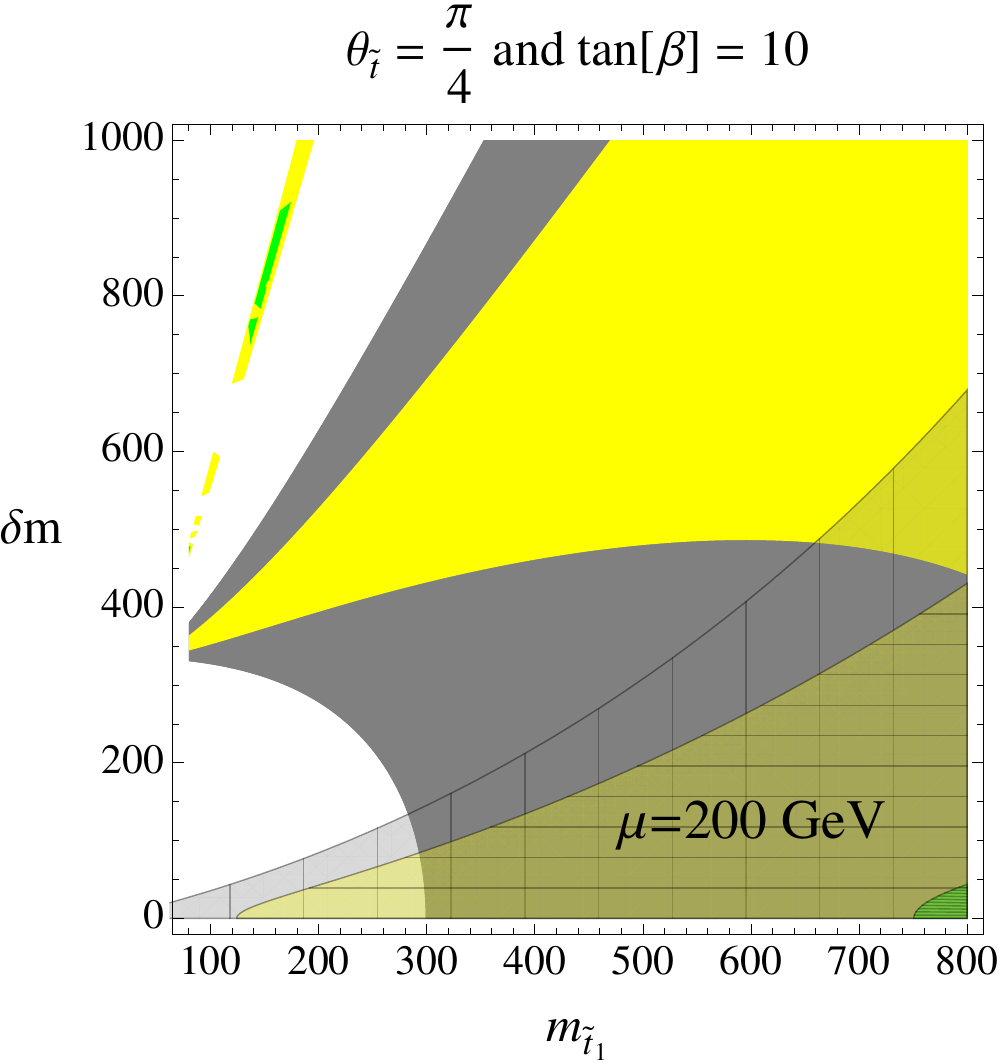}
\caption{Overlay of the best-fit regions of stop parameter space from global fit to Higgs signal strength data with the allowed space due to
${\rm Br}(\bar{B} \rightarrow X_s \, \gamma)$ at the level of $1,2,3 \,\sigma$. The $3 \, \sigma$ allowed region is overlaid with a vertical mesh,
the $2 \, \sigma$ has a horizontal and vertical mesh, while the $1 \, \sigma$ allowed region is the green region with the further addition of the diagonal mesh. In V1 an V2 of this paper a sign error in the Appendix loop functions affected these results allowing a spike region for large $\mu$ to be consistent with ${\rm Br}(\bar{B} \rightarrow X_s \, \gamma)$ constraints.}
\label{btos}
\end{figure}

We use the results of Ref.~\cite{Grzadkowski:2008mf} for the constraints on the BSM Wilson coefficients $C_i = C_i^{SM} + \Delta C_i$
set by the observable ${\rm BR}(\bar{B} \rightarrow X_s \, \gamma)_{E_\gamma > 1.6 \, {\rm GeV}}$. The contribution of the BSM Wilson coefficients to this observable is given by
\bea
{\rm BR}(\bar{B} \rightarrow X_s \, \gamma)_{E_\gamma > 1.6 \, {\rm GeV}} = \left[( 3.15 \pm 0. 23 ) -8.0 \, \Delta C_7(\mu_0)  -1.9 \, \Delta C_8(\mu_0)\right] \times 10^{-4}\ .
\eea
Here we neglect $(\Delta C_i)^2$ terms and have used (implicitly) the input values listed in Ref.~\cite{Grzadkowski:2008mf}, where, in particular, the scale $\mu_0 = 160 \, {\rm GeV}$ was chosen 
for the SM results. Another assumption used in Ref.~\cite{Grzadkowski:2008mf} is that all other induced BSM operators have Wilson coefficients
that satisfy
\bea
\frac{C(\mu \sim \mu_0)}{G_F \, \Lambda} \sim \mathcal{O}(g_W^n),  \quad n \geq 2,
\eea
and the NSUSY scenarios we are interested in will have to satisfy this condition when this constraint is strictly applied. Comparing to the current world experimental (HFAG) average given in Ref.~\cite{Asner:2010qj}
\bea
{\rm BR}(\bar{B} \rightarrow X_s \, \gamma)_{E_\gamma > 1.6 \, {\rm GeV}} = \left[3.43 \pm 0.21 \pm 0.07 \right] \times 10^{-4},
\eea
we will use the $1 (2) \, \sigma$ bound $ -8.0 \, \Delta C_7  -1.9 \, \Delta C_8 = 0.28 \pm 0.32 (0.64)$
to constrain the NSUSY parameter space.\footnote{We have updated our numerical results in V3 of this paper
from the result in Ref.~\cite{Asner:2010qj} to include the new HFAG averaged result that includes the recent Babar result, ${\rm BR}(\bar{B} \rightarrow X_s \, \gamma)_{E_\gamma > 1.6 \, {\rm GeV}} = \left[3.31 \pm 0.16 \pm 0.30 \pm 0.10 \right] \times 10^{-4}$, Ref.~\cite{:2012iw}, with a lower central value but larger error.}  
In Fig.~\ref{btos} we show the interplay of the constraints from the global fit to Higgs data and constraints due to ${\rm Br}(\bar{B} \rightarrow X_s \, \gamma)$  in minimal NSUSY. We see that there exists consistent parameter space that can pass both experimental tests, primarily 
at the level of $\sim 2 \sigma$ in each case. Large mass splittings scenarios of the stop states when large mixing is present are significantly disfavoured.
\subsection{Electroweak Precision Data} \label{EWPD}

Measurements of $m_W$ and other EW precision observables also restrict the allowed parameter space of the NSUSY scenario. 
Recent measurements of $m_W$ at the Tevatron \cite{Aaltonen:2012bp,Abazov:2012bv} are of particular interest. The world average \cite{TevatronElectroweakWorkingGroup:2012gb} has been refined
to $(m_W)_{exp} = 80.385 \, \pm 0.015 \, {\rm GeV}$, with a significant reduction of the quoted error. As recently re-emphasized in Refs.~\cite{Lee:2012sy,Barger:2012hr}, precise measurements of the value of $m_W$ 
constrain the allowed parameter space of a weak scale NSUSY spectra when $m_h$ is known. This occurs as the allowed custodial symmetry violation that could be present in the sfermion sector is bounded.
Note that a global fit to EWPD produces a $\Delta T$ constraint that has about twice the error of the constraint used here, which is directly determined from the shift in the $W$ mass. 

In this Section, we add this further constraint in the study of the impact of  NSUSY spectra on Higgs properties. 
We use the numerical approximation of the two-loop SM prediction of $m_W$ given by Ref.~\cite{Awramik:2003rn} and the method of Ref.~\cite{Barger:2012hr} .
The relevant SUSY correction to the SM prediction\footnote{Here $s_W$ is defined in the on-shell scheme.} of $m_W$ is given in Refs.~\cite{Barbieri:1983wy,Drees:1990dx,Chankowski:1993eu,Heinemeyer:2006px} as
\bea
(\Delta m_W)^{SUSY} \simeq \frac{m_W \, c_W^2}{2 \, (c_W^2 - s_W^2)} \, \Delta \rho^{SUSY}.
\eea
Neglecting terms proportional to small $\tilde{b}$ mixing angles, the SUSY contribution is given by
\begin{align}\label{deltarho}
\Delta \rho_0^{SUSY} &\simeq  \frac{3\, G_F }{8 \, \sqrt{2} \, \pi^2} \, \left\{
\sum_{i=1,2} |\langle \tilde{t}_L|\tilde{t}_i\rangle|^2\,F_0[m^2_{\tilde{t}_i}, m^2_{\tilde{b}_L}]
- |\langle \tilde{t}_L|\tilde{t}_1\rangle|^2|\langle \tilde{t}_L|\tilde{t}_2\rangle|^2\, F_0[m^2_{\tilde{t}_1}, m^2_{\tilde{t}_2}] \,  \right\},
\end{align}
and the function $F_0$ is defined as
\bea
F_0[x, y] = x + y - \frac{2 \, x \, y}{x - y} \, \log \frac{x}{y}.
\eea
The subscripts $0$ are to remind the reader that this is simply the contribution of the third generation sfermions to $\Delta \rho^{SUSY}$.
We can trade $m_{\tilde{b}}$ in this expression for the stop masses and the stop mixing angle ($\theta_{\tilde{t}}$) using 
\bea
m_{\tilde{b}_1}^2 \simeq  \cos^2 \theta_{\tilde{t}} \, m_{\tilde{t}_1}^2 + \sin^2 \theta_{\tilde{t}} \, m_{\tilde{t}_2}^2 - m_t^2 - m_W^2 \, \cos (2 \, \beta).
\label{softSU2}
\eea
This assumes small sbottom mixing and neglects perturbative corrections. 
We are also neglecting other SUSY contributions to $\Delta \rho^{SUSY}$, namely the $\chi^{\pm,0}$ contributions from the light states retained in the residual NSUSY spectrum.
The masses of these states are set by the same scale ($\sim \mu$) in NSUSY,  and the mass splitting of these doublets
is small, leading to subdominant contributions to  $\Delta \rho^{SUSY}$.
\begin{figure}[t]
\centering
\includegraphics[scale=0.58]{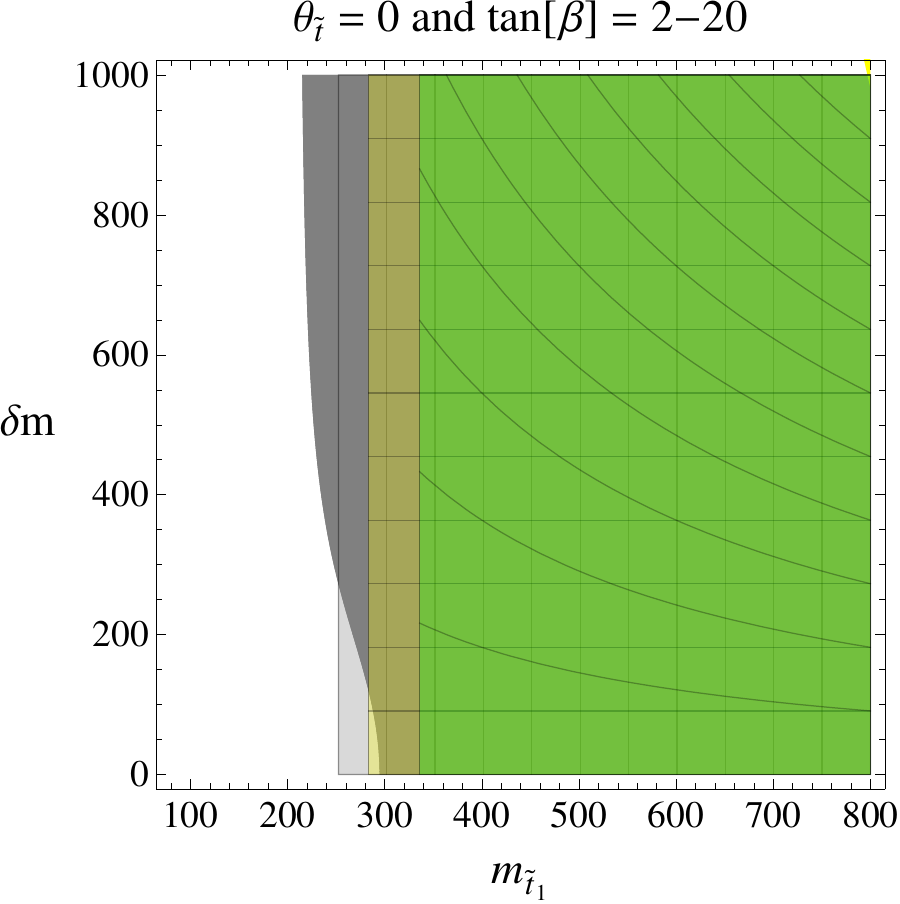}
\includegraphics[scale=0.58]{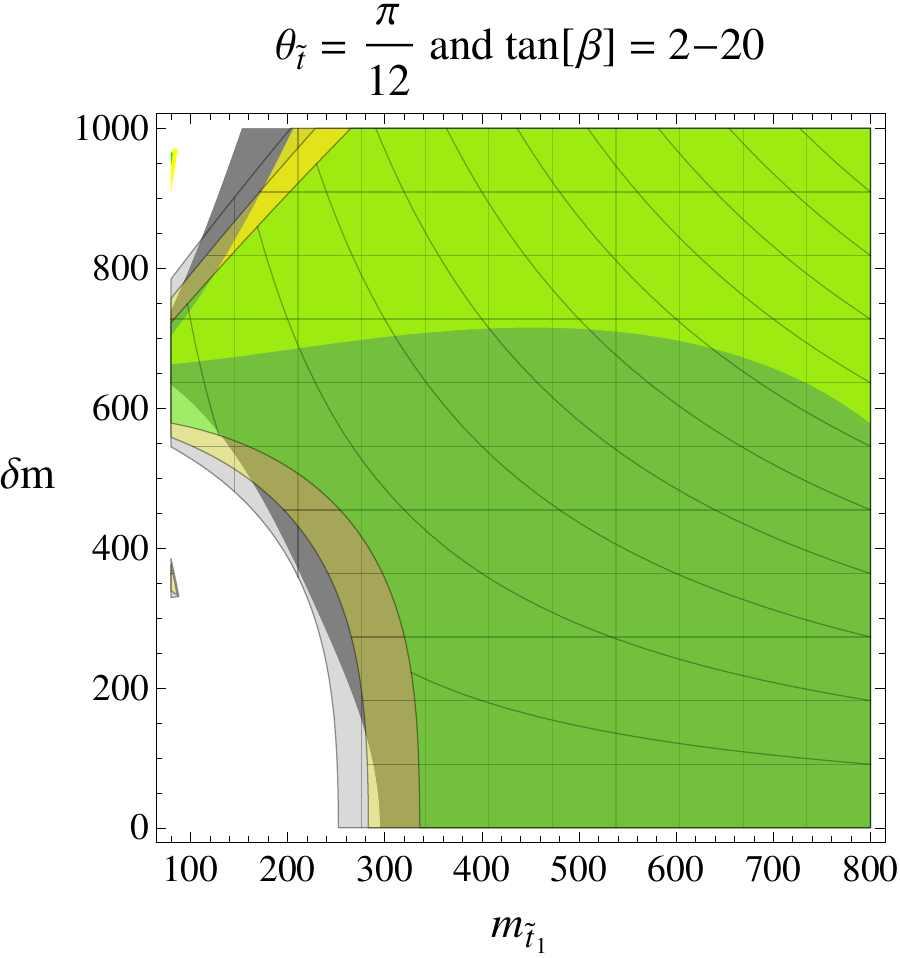}
\includegraphics[scale=0.58]{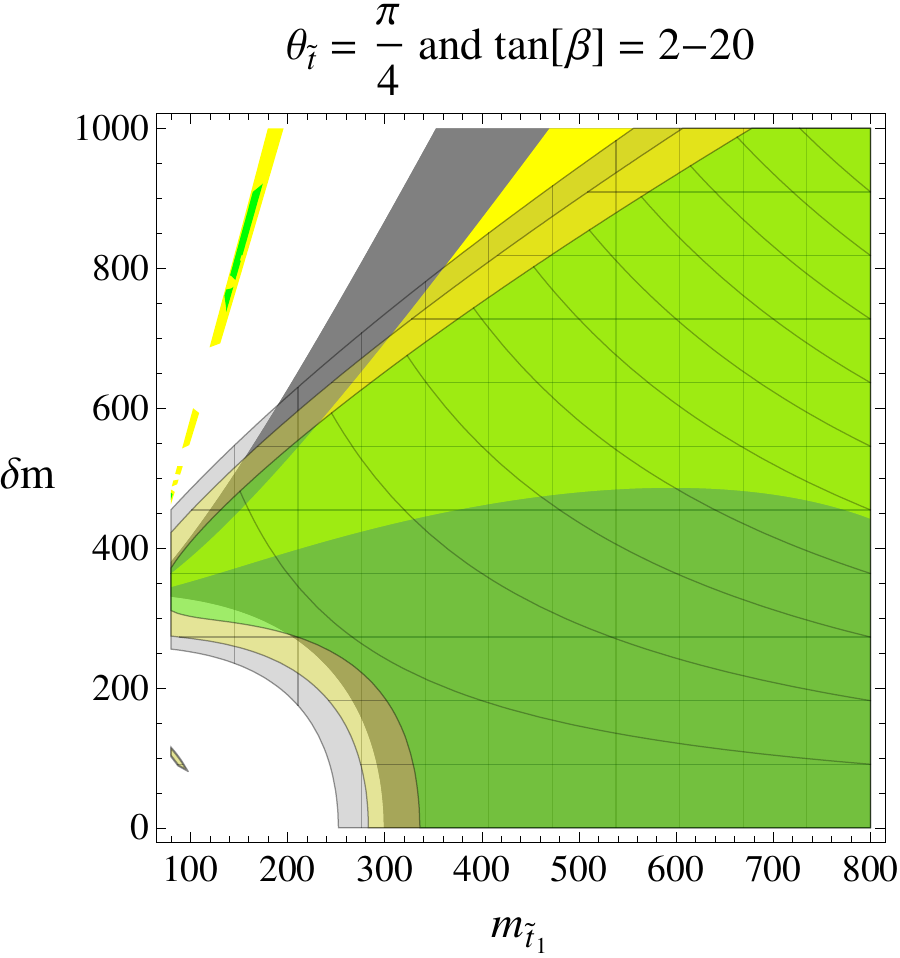}
\caption{Constraints from precision measurements of $m_W$, shown as $1,2,3$ sigma shaded regions, overlaid on the allowed stop parameter space consistent with the global fit to Higgs signal strength data. Plot colour/display convention is the same as in previous figures.}
\label{mw}
\end{figure}
In order to define the SM prediction of $m_W$ we must specify a central value and error in $m_h$.
Given the current state of the data
we simply take the Higgs mass range $125 \pm 2  \,{\rm GeV}$ for this constraint.
The effect of this uncertainty in $m_h$ is very subdominant to the largest source of error, 
which is the uncertainty in $m_t$ (which leads to a $\pm 5.4 \, {\rm MeV}$ effect on $m_W$).

With the choice of input parameters in Table \ref{table:table2}, we find $(m_W)_{SM} = 80.368 \, \pm 0.006 \, {\rm GeV}$, which constrains NSUSY through the resulting bound
$(\Delta m_W)^{SUSY} \lesssim 0.017 \pm 0.016 \, {\rm GeV}$. 
This translates into a constraint $\Delta \rho^{SUSY} \lesssim \left(3.0 \pm 2.8 \right)\times 10^{-4}$.
This is in good agreement with the result of Ref.~\cite{Barger:2012hr}, which uses the same method, up to small differences in the input parameters.

Note that for $\theta_{\tilde{t}}> \pi/4$ the lightest stop is dominantly $\tilde{t}_R$  compared to the 
interval we discuss, $0 < \theta_{\tilde{t}}< \pi/4$, in which it is dominantly $\tilde{t}_L$. However, in the former case one still obtains a similar constraint space for the $\Delta m_W$ constraint, with the shift in the SM prediction and the best fit value selecting for degenerate stop states.

\subsubsection{The Funnel Region}
We show the overlap of the global Higgs fit constraints and the constraints due to $m_W$ in Fig.~\ref{mw}.
Note the good degree of consistency between both constraints. This consistency can be traced to the following. The low mass ``funnel region" in stop parameter space arises from minimizing the 
contribution to $F_g$. The condition $F_g \rightarrow 0$  translates into the relationship
\bea
\label{fgcond}
\sin (2 \, \theta_{\tilde{t}}) \simeq \frac{2 \, m_t \, (m_{\tilde{t}_1}^2 + m_{\tilde{t}_2}^2)^{1/2}}{(m_{\tilde{t}_2}^2 - m_{\tilde{t}_1}^2)}.
\eea
When the stop and sbottom masses are not widely split, it is an excellent approximation \cite{BHR} to use
$F_0[x,y]\simeq (4/3)(\sqrt{x}-\sqrt{y})^2$, which leads to 
\bea
\Delta \rho_0^{SUSY}  \simeq  \frac{G_F }{2 \, \sqrt{2} \, \pi^2}
\left(m_{\tilde{t}_1} \cos^2\theta_{\tilde{t}}+m_{\tilde{t}_2} \sin^2\theta_{\tilde{t}}-
m_{\tilde{b}_1} \right)^2\ ,
\eea
showing that  $\Delta \rho_0^{SUSY} $ has a minimum near
\be
\label{rhocond}
m_{\tilde{t}_1} \cos^2\theta_{\tilde{t}}+m_{\tilde{t}_2} \sin^2\theta_{\tilde{t}}\sim m_{\tilde{b}_1}\ .
\ee
At that point the approximation for $\Delta \rho_0^{SUSY}  $ written above gives exactly zero.
Near the minimum, the result for $\Delta \rho_0^{SUSY}  $ is non-zero
but suppressed and of order 
\bea
\Delta \rho_0^{SUSY}  \simeq  \frac{G_F }{8 \, \sqrt{2} \, \pi^2} \left(\frac{m_t^4}{m_{\tilde{t}_2}^2} \right),
\eea
(for $m_{\tilde{t}_2} \gg m_{\tilde{t}_1},m_t$) which is of the right order of magnitude to match the condition $\Delta \rho^{SUSY} \sim 10^{-4}$. 
Note that away from the minimum $\Delta \rho_0^{SUSY}$ could be larger by a factor $m_{\tilde{t}_2}^2/m_{\tilde{t}_1}^2$ (always in the limit $m_{\tilde{t}_2} \gg m_{\tilde{t}_1}$) which would destroy the compatibility with the measurement of $m_W$.

The condition (\ref{rhocond}) can be rewritten as
\be
\label{rhocond2}
\sin^2 (2 \, \theta_{\tilde{t}}) \simeq \frac{4 \, m_t^2}{(m_{\tilde{t}_2} - m_{\tilde{t}_1})^2},
\ee
which coincides with  (\ref{fgcond}) up to a factor $(m_{\tilde{t}_1}+m_{\tilde{t}_2})^2/(m_{\tilde{t}_1}^2+m_{\tilde{t}_2}^2)$, which is $\sim 1$ for $m_{\tilde{t}_2}\gg m_{\tilde{t}_1}$. This slight mismatch between the conditions (\ref{fgcond}) and (\ref{rhocond}) gives rise to a non-zero $F_g$:
\bea
F_g \simeq \frac{2 m_t^2}{3  m_{\tilde{t}_1} m_{\tilde{t}_2}},
\eea
which can nonetheless remain small enough to fit the LHC Higgs data.
In this sense, this funnel region is clearly associated with some cancelations due to parameter tuning.

\subsection{Collider Bounds}\label{collider-bounds}

In this Section we describe some of the current collider bounds on the NSUSY spectrum and how they pertain to the stop parameter space of interest found in previous sections.
The most studied and stringent bounds on NSUSY come from missing energy signatures. However, including R-parity or not in a weak scale NSUSY spectrum is not dictated directly by naturalness.
Due to this, we will mostly restrict our attention to more robust collider constraints. We note however that with the absence (to date) of missing energy signatures,
several searches for R-parity violating phenomenology (without significant missing energy)~\cite{atlas-cms-general} in the multijet and multilepton final states are now ongoing.
These studies could further constraint the interesting parameter space that we isolate in this study. 

The main features of NSUSY relate to the Higgsino and Stop sectors. As we have discussed before, if Higgsinos get their mass solely from the $\mu$ term, we expect a very small mass splitting between the $\chi^\pm,\chi^0$, of the order of ${\cal O} ( v \sqrt{\mu/M_{1,2}})$. Here $M_{1,2}$ are the gaugino masses that are taken to be large $M_{1,2} \gg v$ in minimal NSUSY.  In the deep-Higgsino region, the $\chi^\pm \rightarrow \chi^0 + X$ decay occurs with no hard-p$_T$ objects to tag on. Then, only monojet/monophoton searches at LEP were sensitive to such decays, leading to a bound $m_{\tilde{\chi}^{\pm}}>$ 92 GeV~\cite{pdg} .

Searches for stops are limited by the complexity of the final state and the similarity with the SM $t\bar{t}$ background, especially in a scenario where the chargino and neutralino are heavy. The recent direct stop combination searches from 
ATLAS~\cite{direct-stop} do not have sensitivity on the stop mass, except in a very small region around $m_{\tilde{t}}\sim$ 300 GeV, and only for a limited range of $\mu$.~\footnote{Note that there are some proposals to improve the stop searches~\cite{Plehn:2012pr,Cao:2012rz,Bai:2012gs,Han:2012fw,Alves:2012ft,Kaplan:2012gd}, using different strategies,  from boosted tops to fitting the missing energy distribution.}
One can obtain more information on the stop sector by looking at searches of a single photon plus missing energy~\cite{CMSmonophoton,atlas-monoa}. Although those searches were performed in the context of large extra-dimensions and dark matter effective theories, a straightforward re-interpretation in terms of SUSY can be done~\cite{monophoton}, within the assumption that the stop and the neutralino are separated by $\leq 30$ GeV. This is indeed the case in a significant amount of the low-mass region preferred by the Higgs fit, namely for $m_{\tilde{t}}\leq$ 120 GeV. Since the Higgsino is heavier than 90 GeV, the splitting between $\tilde{\chi}$ and $\tilde{t}$ is right at the best sensitivity point.  
The mass bound obtained in Ref.~\cite{monophoton} is 150 GeV (95\% C.L.). This strongly constrains/excludes some of the low mass stop parameter space most preferred in the global Higgs fits. Monojet searches could also be used in the low $m_{\tilde{t}_1}$ region of parameter space~\cite{monojet-baer}, and one could also use top precision measurements (spin correlations) to potentially rule out this area of parameter space~\cite{Han:2012fw}.

Finally, we note that gluino-assisted stop production searches rely on a light gluino around a TeV, which is in some tension with the combination of multijets+leptons+missing energy~\cite{atlas-multijet,new-cms-multijet}, which leads to a bound of 1050 GeV. This gluino bound is rather model independent, as intermediate $\tilde{\chi}^{\pm}$, $\tilde{\chi}_2$ to leptons or off-shell stops do not change the bound considerably~\cite{PAS-SUS-11-016,PAS-SUS-12-016}.\footnote{Although this search is optimized for Majorana gluinos, the limit would be even more stringent for Dirac gluinos~\cite{N2}.} Nevertheless, a gluino heavy enough to disable this search is allowed in a minimal NSUSY scenario and the
stop limits from these gluino-assisted studies do not (as yet) directly constrain the parameter space of interest.

\section{Combining Constraints}\label{combined}

It is of interest to address how a NSUSY scenario globally fits the increasingly rich dataset. In past sections, we have shown the interplay of various precision constraints;
in this Section we perform a global $\chi^2$ fit to Higgs signal-strength data, constraints due to precision measurements of $m_W$
and ${\rm Br}(\bar{B} \rightarrow X_s \, \gamma)$. We fix $\tan \beta = 10$ 
and show in Fig.~\ref{globalfit} such global fit results for the three mixing cases we have considered
in this paper,  to illustrate the allowed global fit space.

\begin{figure}[t]
\centering
\includegraphics[scale=0.58]{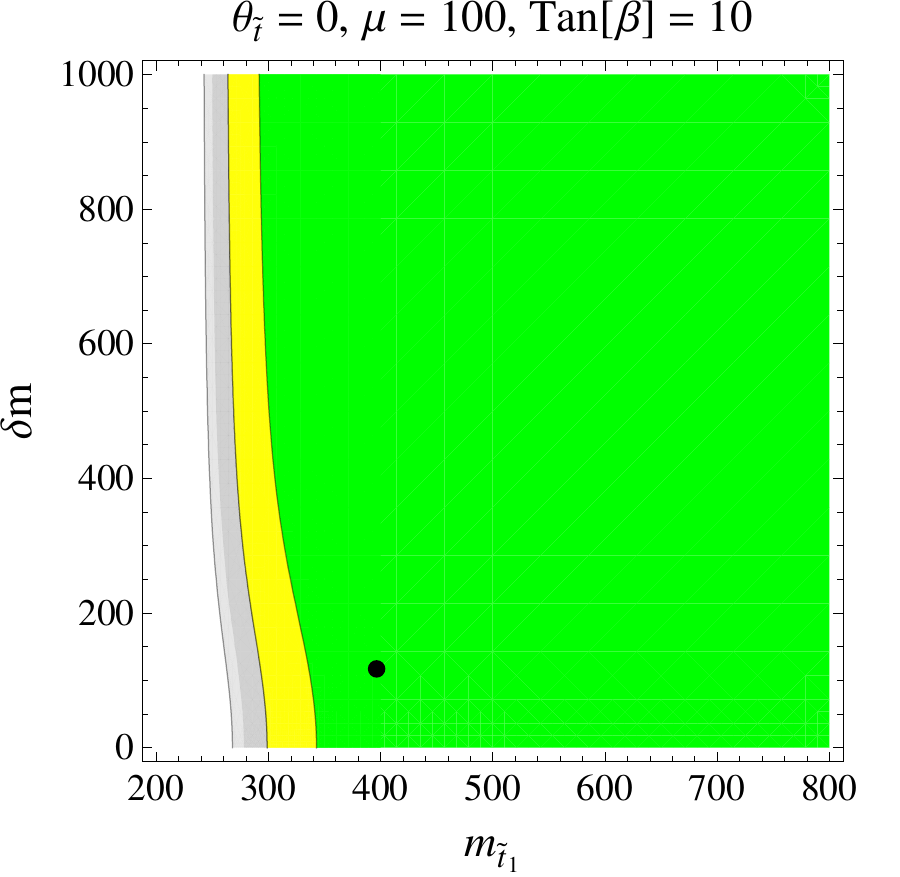}
\includegraphics[scale=0.58]{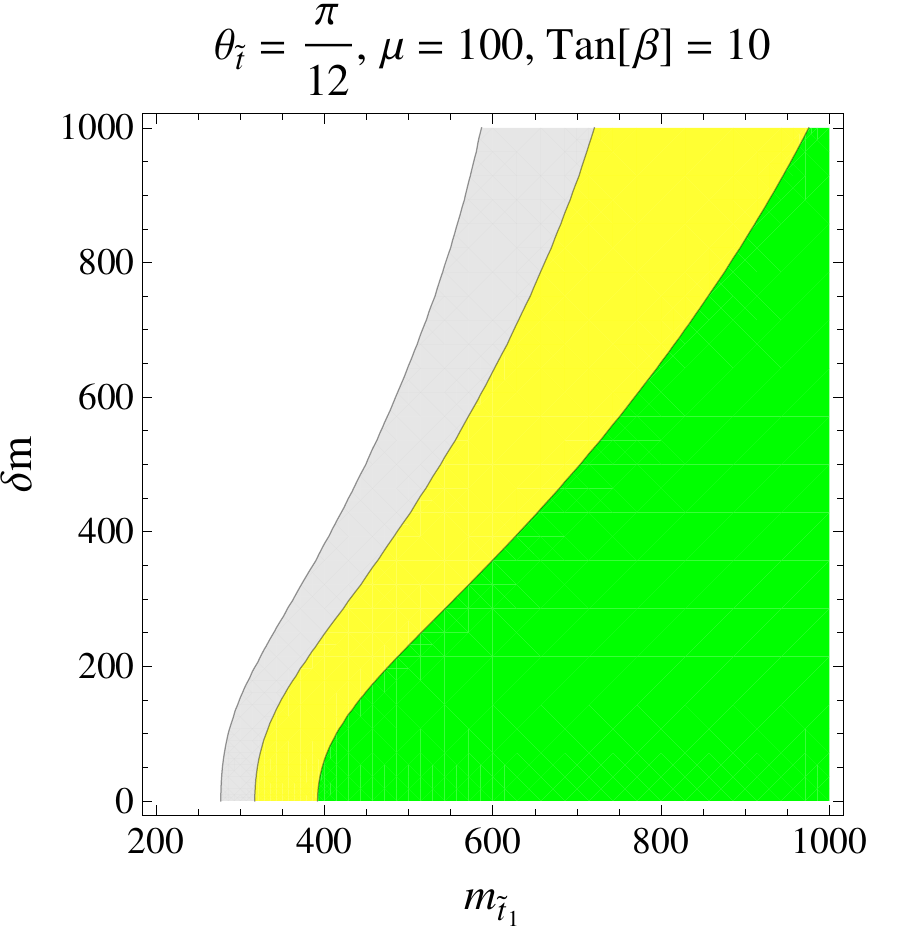}
\includegraphics[scale=0.58]{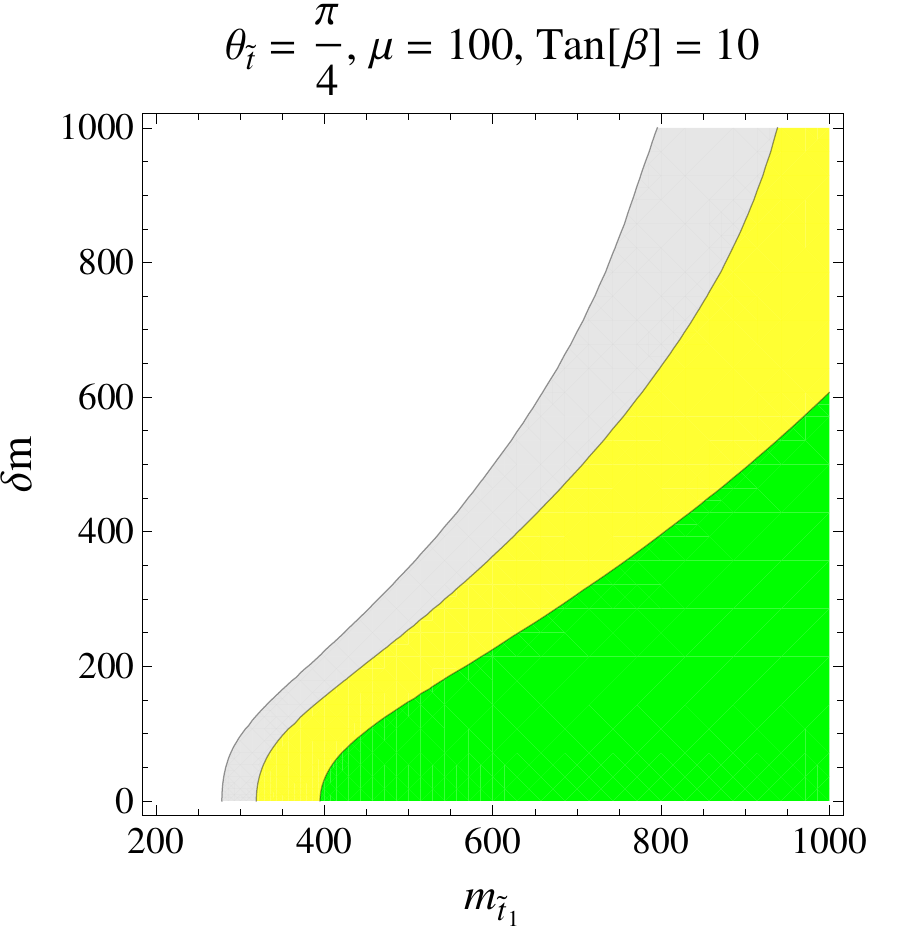}
\includegraphics[scale=0.58]{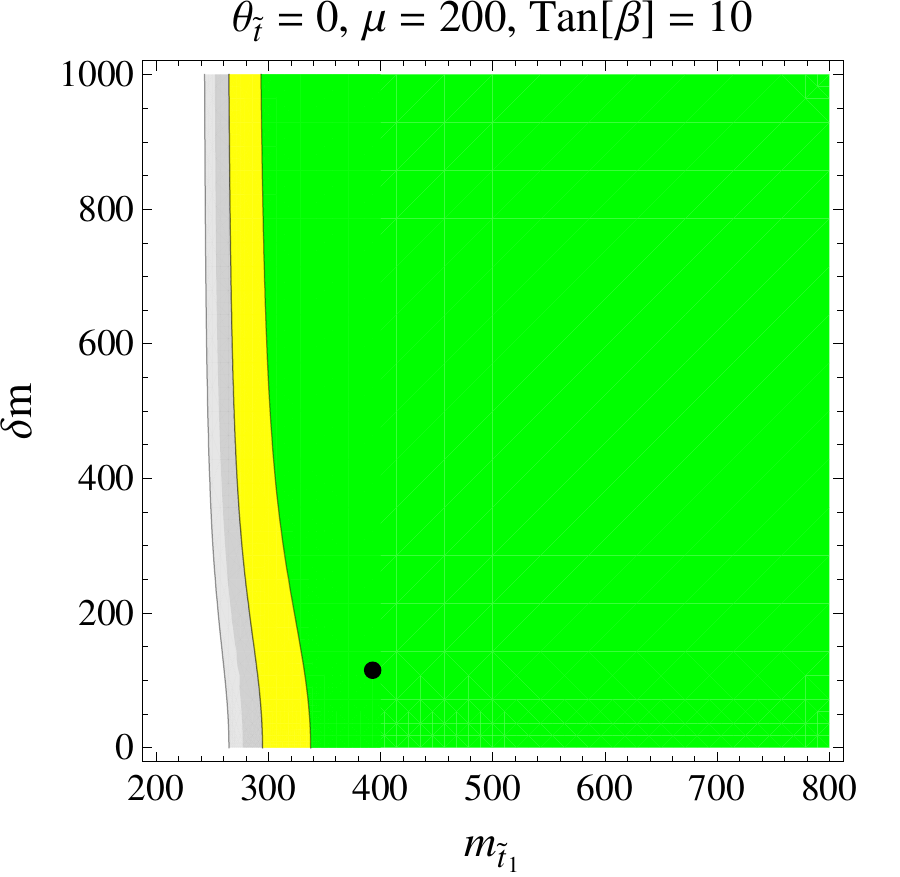}
\includegraphics[scale=0.58]{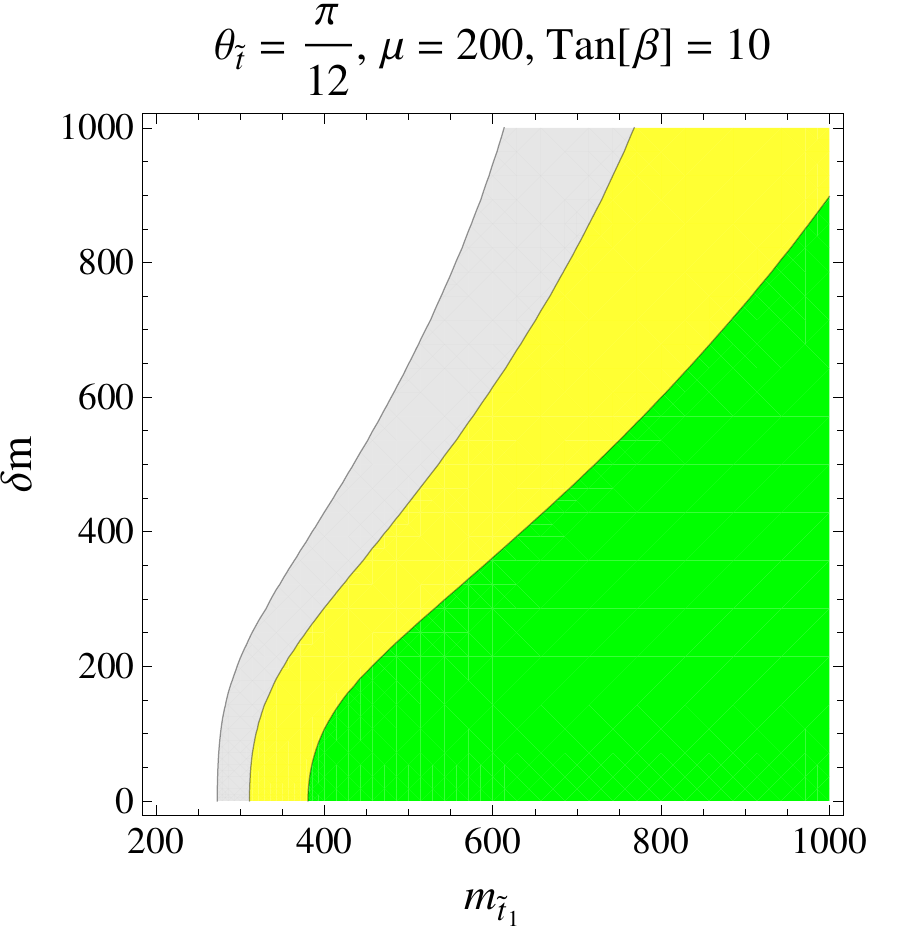}
\includegraphics[scale=0.58]{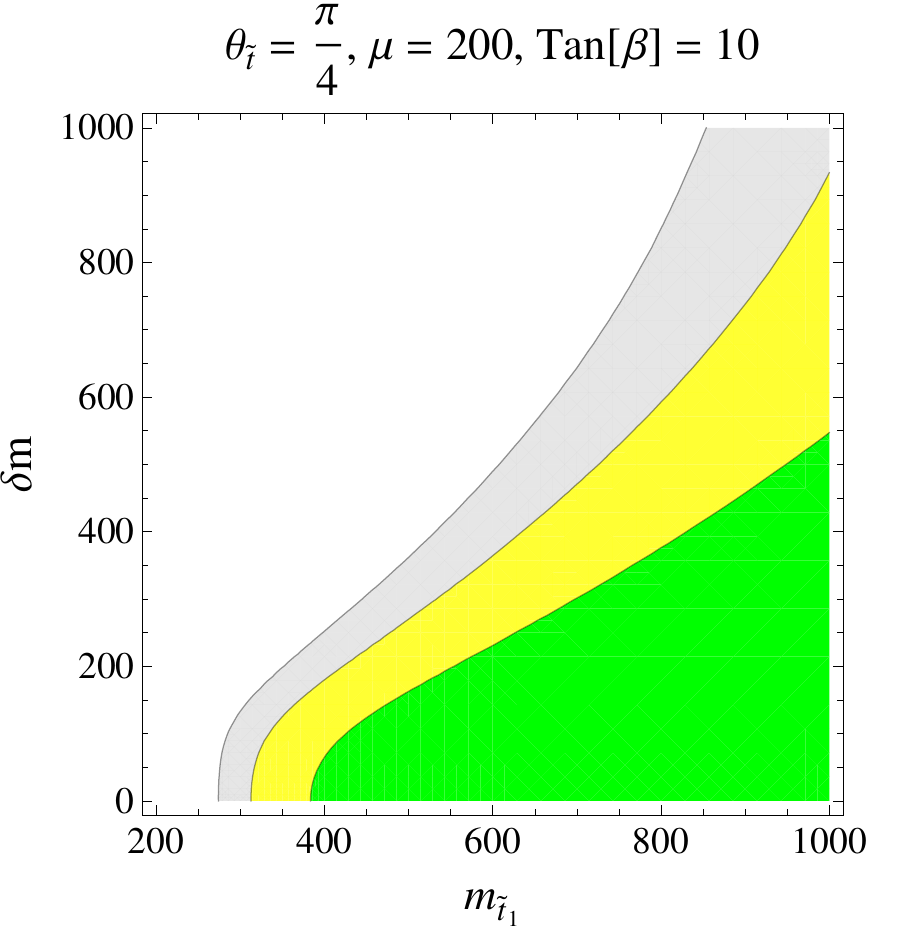}
\caption{Best-fit regions in stop parameter space from the global $\chi^2$. Colour conventions the same as in previous figures. We show the three mixing cases $\theta_{\tilde{t}} = 0, \pi/12, \pi/4$ for two values of $\mu$ that set the Higgsino mass scale
in these minimal scenarios, $\mu = 100,200 \, {\rm GeV}$. The $\tan \beta$ dependence of the result is primarily driven by the ${\rm Br}(\bar{B} \rightarrow X_s \, \gamma)$ constraint. Larger values
of $\tan \beta$ select for a more degenerate stop spectrum when $\theta_{\tilde{t}} \neq 0$. In 
determining $\chi^2_{min}$, we have assumed a prior $150 \, {\rm GeV} \leq m_{\tilde{t}_1}$ , due to monophoton constraints. In each figure the best fit point is marked with a dot if present in the shown space. 
Note that for the plots where $\theta_{\tilde{t}} \neq 0$, the best fit point is greater than 1 $\rm{TeV}$, and this is why the point is absent. However 
we note again that the $\chi^2$ is a shallow function in the degenerate stop mass best fit region.}
\label{globalfit}
\end{figure}

The results in Fig.~\ref{globalfit} show a good fit to the data. We are fitting to fifty observations: forty-eight Higgs signal-strength measurements,
as well as  $\Delta m_W$ and ${\rm Br}(\bar{B} \rightarrow X_s \, \gamma)$. For the three mixing cases $\theta_{\tilde{t}} = 0, \pi/12, \pi/4$ and $\mu = 100, 200 \, {\rm GeV}$
we find, fitting to the two stop parameters $m_{\tilde{t}_1}$, $\delta m$, that $\chi^2_{min}/(d.o.f.) \simeq 1$.
The interplay of the constraints that allows a good fit is of interest. The best-fit region in the Higgs 
signal-strength fit at very low masses, $m_{\tilde{t}_1} \sim 100 \, {\rm GeV}$, (See Fig.2 (right)) is essentially ruled out due to mono-photon searches and constraints from 
$\Delta m_W$ and ${\rm Br}(\bar{B} \rightarrow X_s \, \gamma)$. When combining constraints,
one finds a best fit to the Higgs signal-strength data with a larger  $m_{\tilde{t}_1}$ (with less mass splitting for larger $\theta_{\tilde{t}}$) in a manner that
also improves agreement with the small deviations from the SM predictions in $\Delta m_W$ for large
regions of parameter space. However, the significance of this observation is currently marginal: the
SM gives a comparable fit to these observables with $\chi^2_{min}/(d.o.f.) \simeq 1.0$.

\subsection{Fine-tuning and the Higgs Mass}

It is also interesting to examine if the Higgs mass could be consistent with its observed value in the allowed parameter space of the fit, without large fine-tuning. We have included the large radiative corrections to the Higgs mass expected after SUSY breaking  \cite{mhrad} by using the public code FeynHiggs \cite{FH0,FH1}
(for alternatives, see \cite{Alt}), which includes all one-loop corrections and the dominant two-loop effects \cite{twoL}. This gives a precise enough determination of $m_h$ (with an error of a few GeV), provided the hierarchy in the stop masses is not so large that further re-summation of logarithms is needed \cite{EN}. Precision is needed here because the soft masses required to give a large enough Higgs mass are exponentially sensitive to $m_h$.
\begin{figure}[h!]
\centering
\includegraphics[scale=0.59]{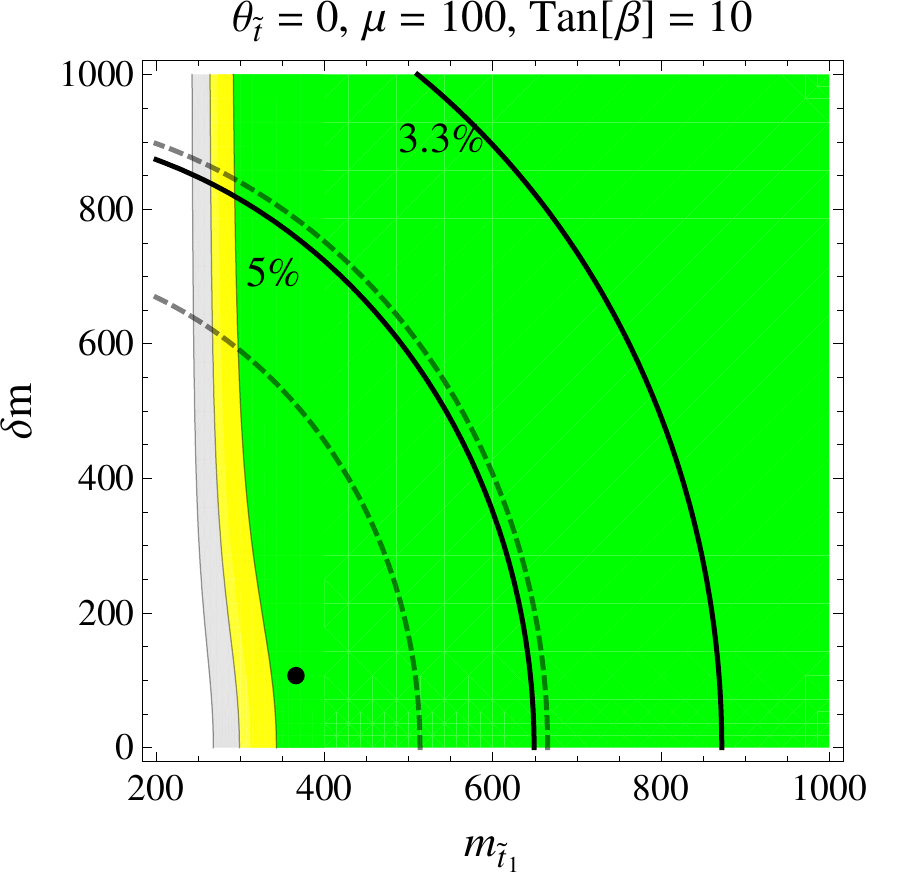}
\includegraphics[scale=0.59]{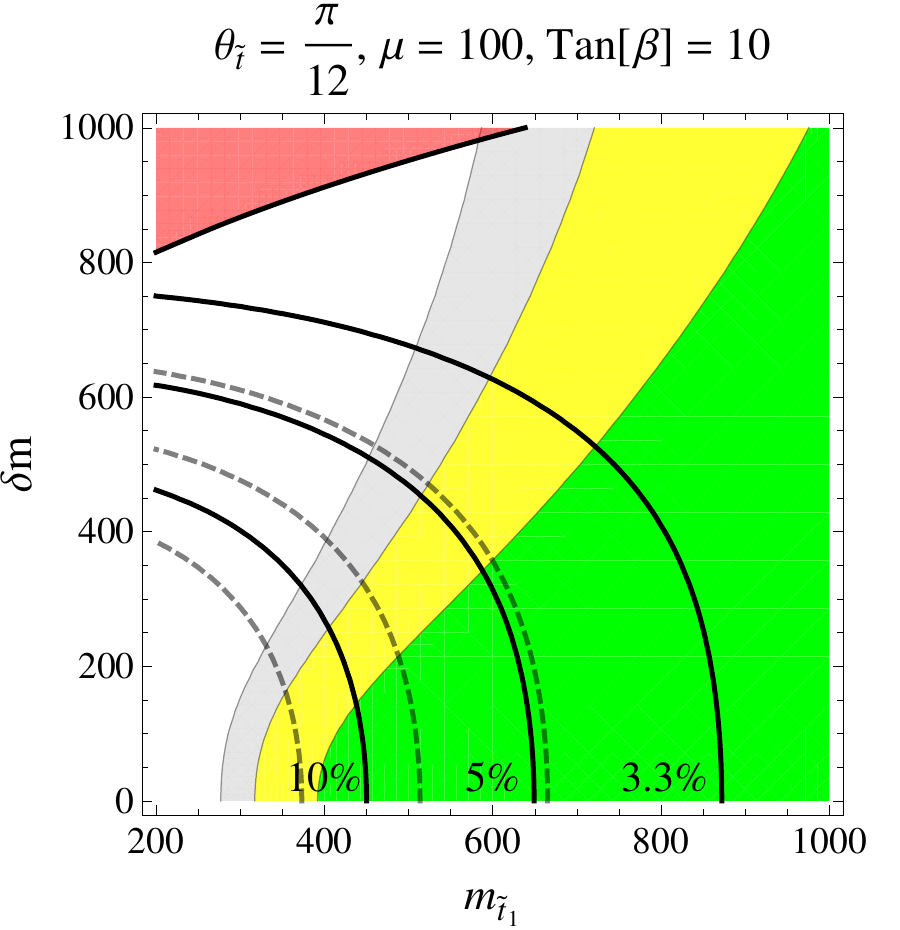}
\includegraphics[scale=0.59]{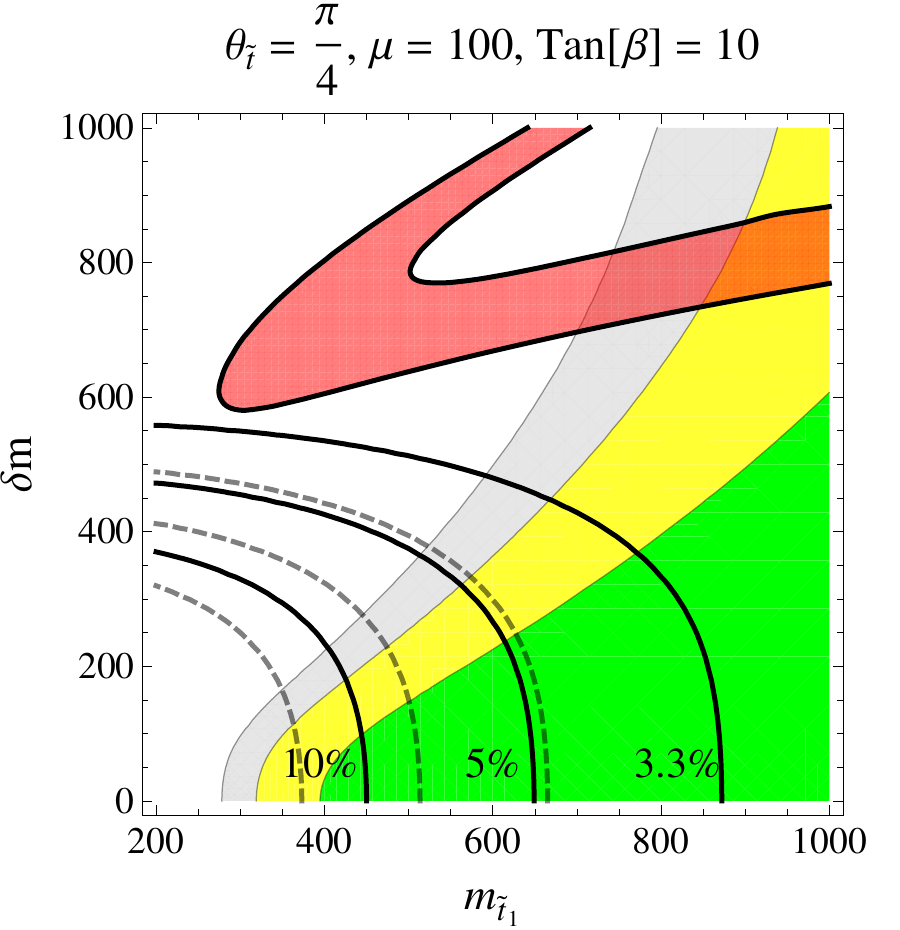}
\includegraphics[scale=0.59]{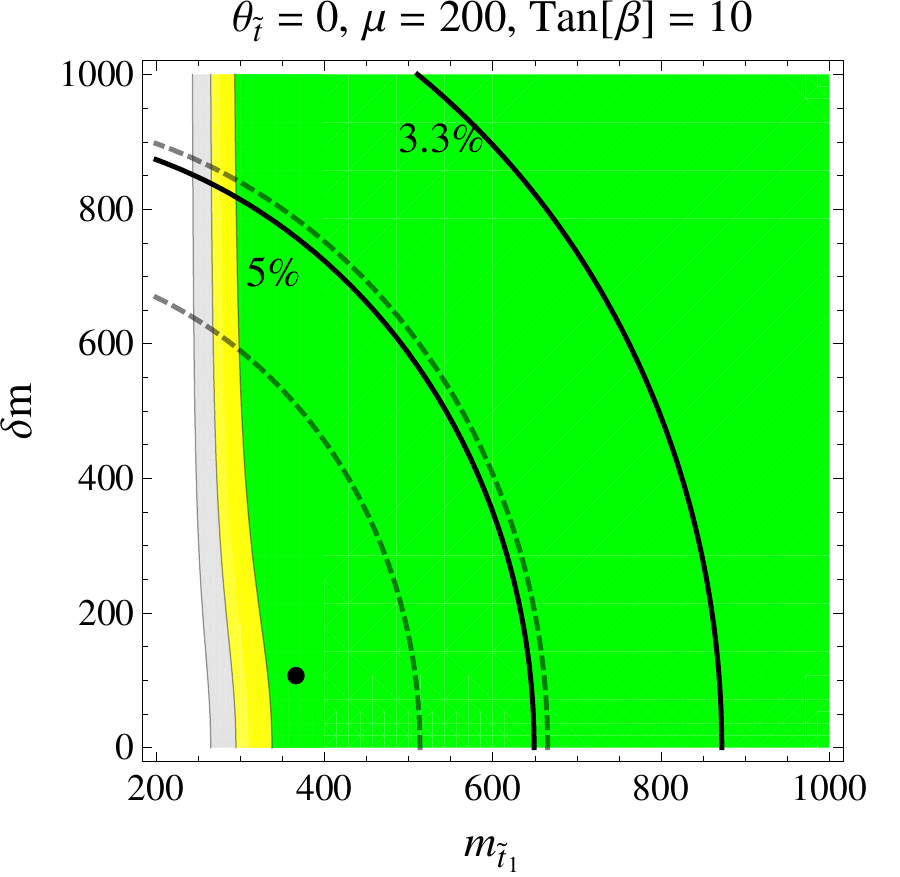}
\includegraphics[scale=0.59]{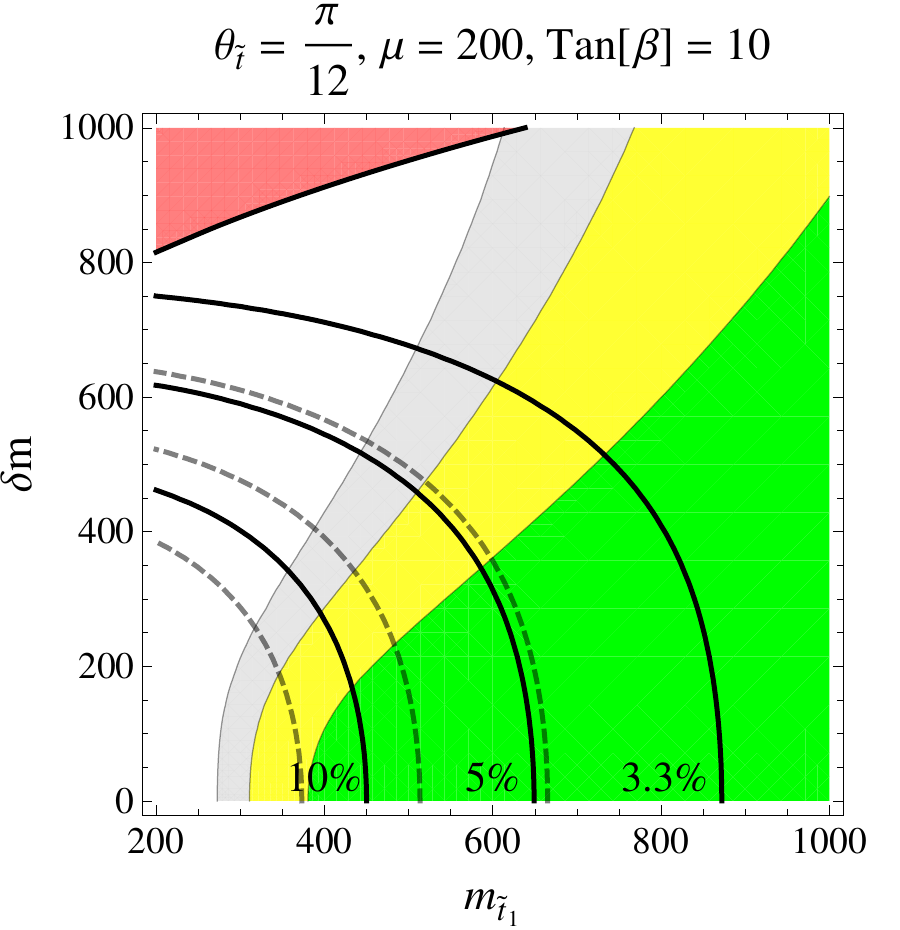}
\includegraphics[scale=0.59]{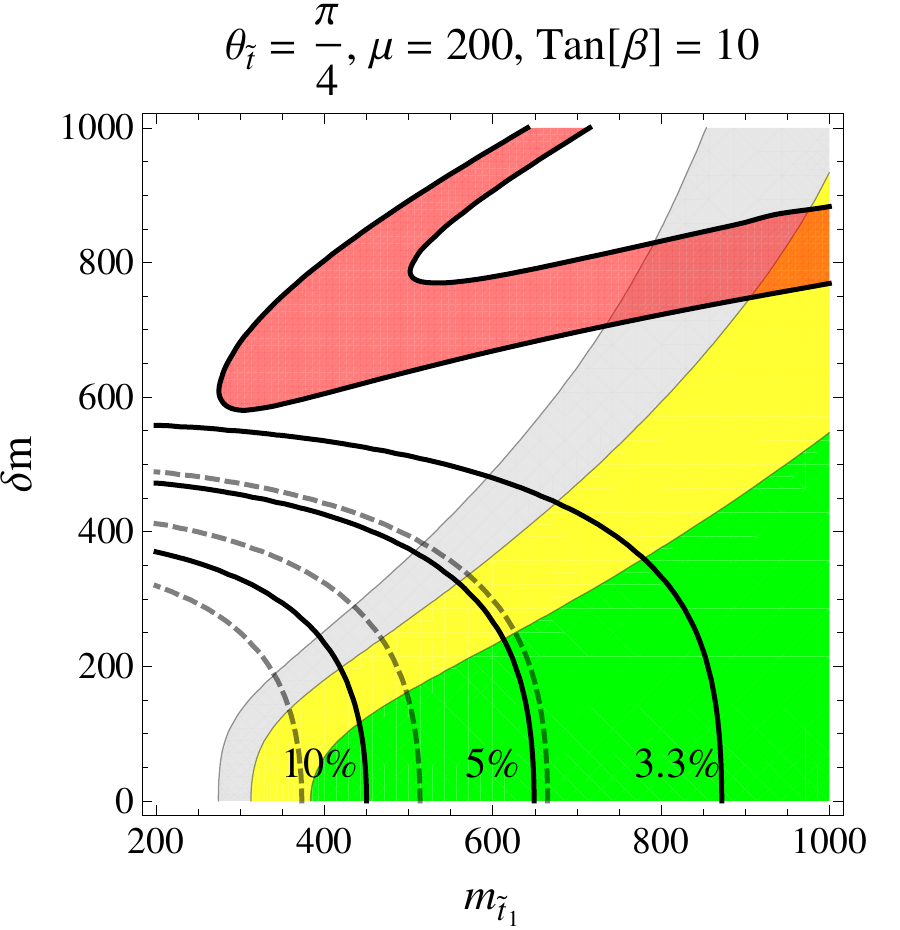}
\caption{The overlay of the Higgs mass condition $m_h=125\pm 2 \, {\rm GeV}$ (in red/darker shaded region) and the best fit space (colour convention the 
same as in previous figures). Also shown are labeled contours of constant fine-tuning (solid black) for a UV cutoff $\Lambda = 10 \, {\rm TeV}$. The same contours are shown (dashed grey) for $\Lambda = 50 \, {\rm TeV}$. In the no-mixing case, the Higgs mass constraint cannot be accommodated in minimal scenarios: no Higgs mass band is present in these plots.}
\label{higgsmass}
\end{figure}

We show in Fig.~\ref{higgsmass} the band in stop parameter space that gives 
$m_h=125 \pm 2 \, {\rm GeV}$, overlaid on the best fit regions. The Higgs mass has a non-negligible
 two-loop sensitivity to the value of the gluino mass, which we have chosen to be 1.1 TeV in this figure.\footnote{We take the other sfermion states to be $\sim 2 \, {\rm TeV}$ in this analysis, however the Higgs mass band is insensitive to the particular mass values of these states for $\tan \beta \sim 10$ and sfermion masses in the ${\rm TeV}$ range.}
For non-zero stop mixing, larger values of the gluino mass tend to lower $m_h$ and therefore require larger values of the stop masses \cite{FH0}. We see in Fig.~\ref{higgsmass}, left plot, that
the best-fit region with sub-TeV stop masses and zero mixing is not consistent with the Higgs mass condition, as was to be expected. Larger stop masses with large mixing can only produce the observed 
Higgs mass value when a worse fit is considered, although the degree of fine-tuning in these parameter regions is a concern. 

It is well known that physics beyond the minimal supersymmetric version of the SM can easily give contributions to the Higgs quartic coupling which controls the Higgs mass. This can happen through new sectors coupled to the Higgs in such a way that they add at tree-level new supersymmetric $F$-term or $D$ term (or even susy-breaking) contributions to the Higgs quartic even in the Higgs decoupling limit (see \cite{wayout1,wayout2,wayout3} for an incomplete list of studies on this). A very moderate tree-level
upward shift of this quartic, which can be interpreted as a threshold correction at the SUSY scale, would have a direct and very important impact on increasing the predicted Higgs mass and reducing the associated fine-tuning.  
The new physics required for this shift can, on the other hand, have a negligible impact on the coupling of the Higgs to gluons or photons. Therefore, in such scenarios one is allowed to soften the link between $m_h$ and the stop sector, while the analysis of the impact of Higgs search results on constraining stop parameters we have performed would still apply. 

It is difficult to define a uniquely compelling fine-tuning measure in an effective theory, or argue what degree of fine-tuning is clearly unacceptable. Fine-tuning considerations are necessarily dependent on the UV physics of the EFT and so one can only make a rough estimate in a low-energy effective theory.  
We could follow the definition of fine-tuning measure of Refs.~\cite{Ellis:1986yg,Barbieri:1987fn}, which 
quantifies the tuning by measuring the sensitivity of the electroweak scale (as given by e.g. the $Z$ mass)
with respect to changes in the fundamental UV parameters. However, as we want to keep an open mind about what detailed
UV physics completes the NSUSY scenario, we will content ourselves with a different
estimate of tuning that simply compares the value of the $Z$ mass with known loop contributions
to it (in our case those coming from stop corrections), requiring that the latter are not much bigger than
$m_Z$. That is, defining
\bea\label{deltaz}
\Delta_Z = \left| \frac{\delta_{\tilde{t}} \, m_Z^2}{m_Z^2} \right|\ ,
\eea
the associated fine-tuning will be 1 part in $\Delta_Z$. Restricting ourselves to the moderate value of $\tan \beta \sim 10$ and assuming that stops loops are the dominant contribution to this fine-tuning measure, one finds (see e.g. Ref.~\cite{Perelstein:2007nx})
\bea
\delta_{\tilde{t}} \, m_Z^2 &\simeq&
\frac{3 m_t^2}{4 \, \pi^2\, v^2} \, \left(M_{\tilde{Q}_L}^2 +
M_{\tilde{t}_R}^2 + A_t^2\right) \, \log \left(\frac{2 \, \Lambda^2}{m_{\tilde{t}_1}^2 + m_{\tilde{t}_2}^2} \right)
\nn\\ 
&\simeq &\frac{3}{8 \, \pi^2\, v^2} \, \left[2m_t^2 (m_{\tilde{t}_1}^2 + m_{\tilde{t}_2}^2 - 2 m_t^2) +
\frac{1}{4} (\delta m)^4  \,\sin^2(2\theta_{\tilde{t}}) \right] \, \log \left(\frac{2 \, \Lambda^2}{m_{\tilde{t}_1}^2 + m_{\tilde{t}_2}^2} \right). \label{finetune}
\eea
The scale $\Lambda$ is associated with new states required to cut off the logarithmic divergence
in the effective theory, and is associated with the messenger scale that transmits SUSY breaking from a hidden sector.
The choice of this scale is UV dependent, offering some further caution on the interpretation of the results. For numerical purposes we consider this scale to be $\Lambda \sim 10,20 \, {\rm TeV}$
consistent with recent choices in the literature \cite{Brust:2011tb}. The fine-tuning contours are overlaid  in Fig.~\ref{higgsmass} in this case (for related work, see Ref.~\cite{Hall}).

Finally, we also note that, if the "funnel region" continues to offer a good fit to the data, such stop masses are very difficult to probe directly in collider searches. A continued deviation consistent with such stop parameters in the global Higgs fits could be the leading experimental indication of the presence of stop states in this exceptional region of parameter space. We discuss the prospects for such future
fits in the next Section.

\subsection{Dark Matter Relic density constraints}
In this Section, we have not imposed any constraints on the spectrum related to achieving the correct Dark Matter (DM) relic density. We briefly note that 
typically, Higgsino dark matter is disfavored. The annihilation rate is too large, leading to a small relic density ~\cite{Jungman:1995df}. In the deep Higgsino region, the chargino is also Higgsino-like and is very close in mass to the neutralino. In this case, coannhilations between charginos and neutralinos become important ~\cite{Griest:1990kh}, as well as coannhihilations with the stops, see Refs.~\cite{Boehm:1999bj,Ellis:2001nx}. Moreover, pure Higgsino DM may be disfavoured by XENON100, see for example a study in the framework of CMSSM ~\cite{Ellis:2012aa,Buchmueller:2012hv}, and also as a possible explanation of FERMI-LAT and PAMELA observations as discussed in Ref.~\cite{Belanger:2012ta}. This situation could be ameliorated if the DM relic density was not entirely due to the neutralino, but there was another relic component, or if the neutralino was not the LSP, but decayed to gravitino, or through RPV interactions. In any case, the focus of this paper is not higgsino DM phenomenology, and we leave a thorough study of the cosmological, indirect and direct detection constraints on pure higgsino dark matter to future work.

\section{Projections for NSUSY at the end of the 8 TeV run}\label{projections}

\subsection{Fit prospects.}

\begin{figure}[t]
\centering
\includegraphics[scale=0.65]{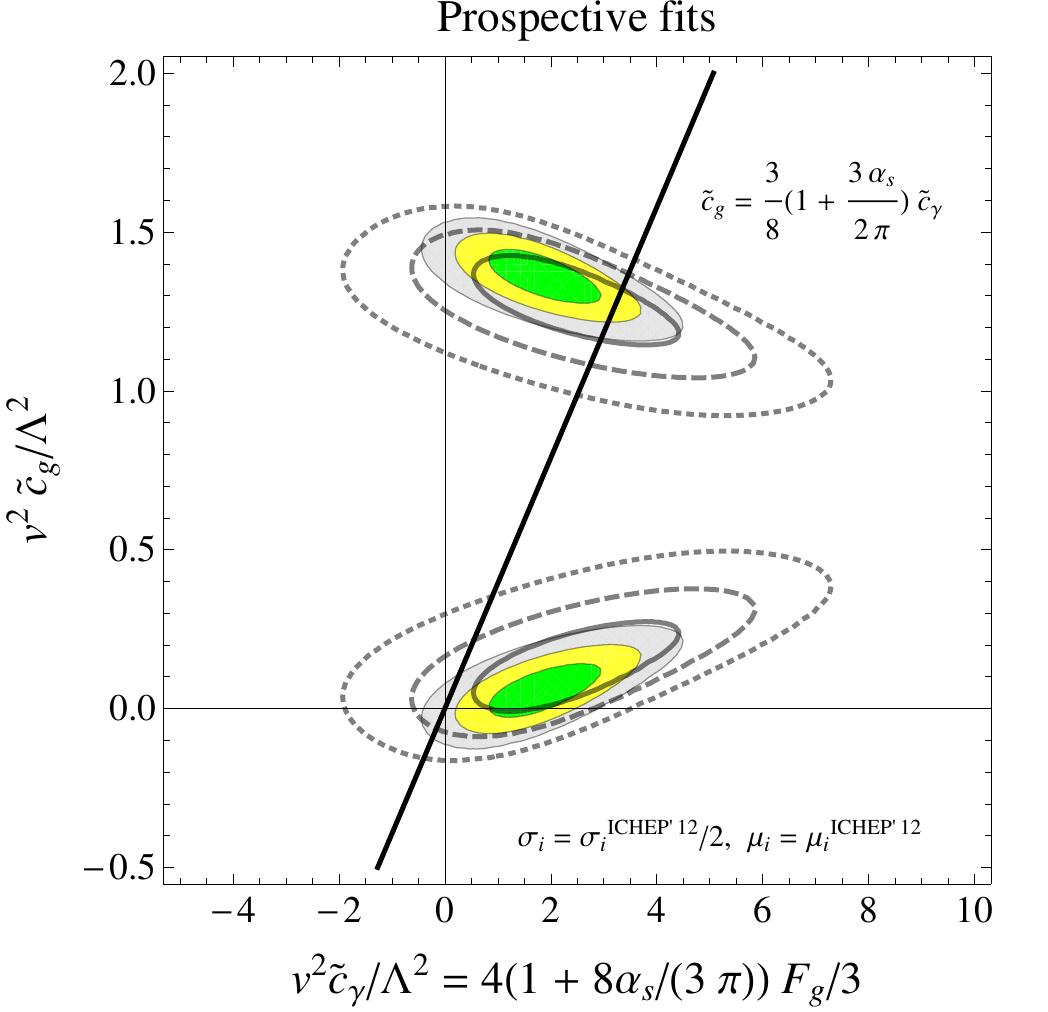}
\includegraphics[scale=0.7]{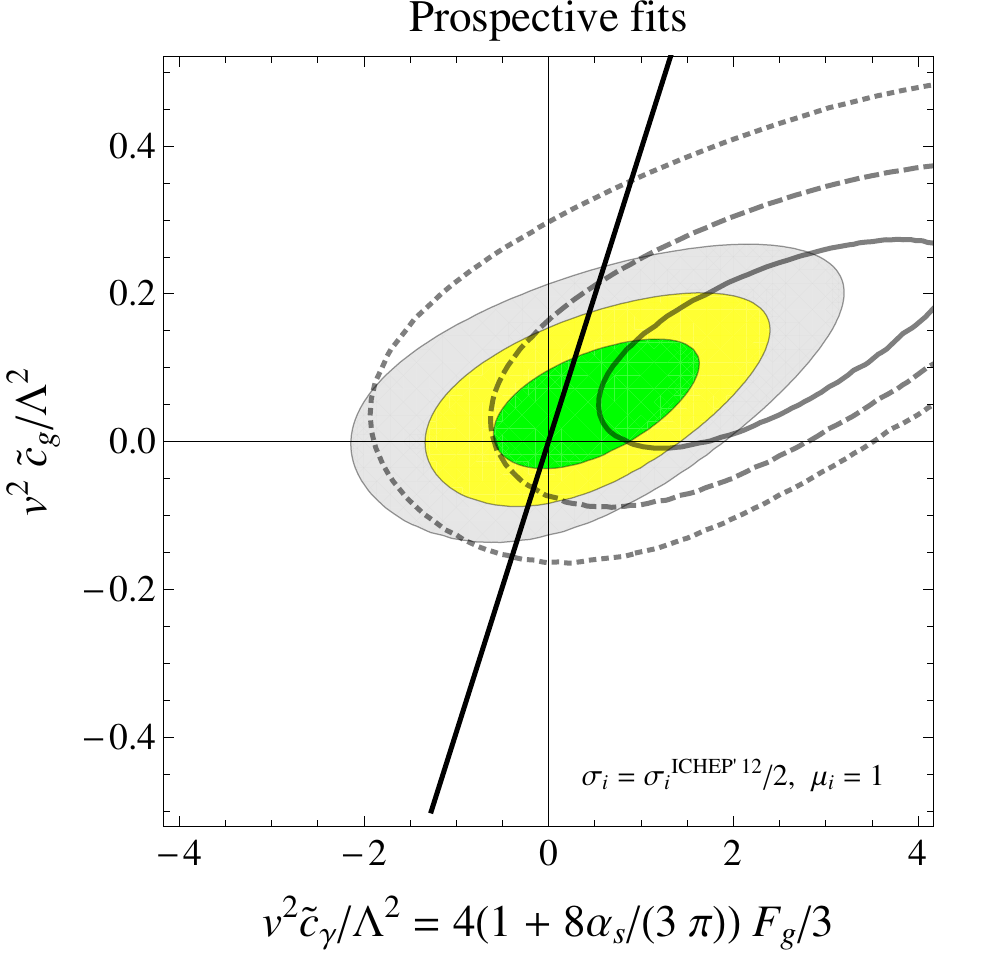}
\caption{Projected best-fit areas when $8 \, {\rm TeV}$ signal-strength values hypothetically report by the end of 2012 have their errors scale down to $\sigma_i/2$ while they keep their current $8 \, {\rm TeV}$ 
best-fit central values (left) and when the best-fit central values converge to the SM (right).
For comparison, we also show in faded gray (solid, dashed, dotted) lines the current $(1,2,3) \, \sigma$ 
best-fit contours.}
\label{prospects}
\end{figure}

The integrated luminosity of experiments at LHC is increasing at a rapid pace, and the power and precision of global Higgs fits of this form have scaled
remarkably to date in the $7,8 \, {\rm TeV}$ runs. If this scaling continues (i.e. if systematics and/or correlations do not become the dominant source of error by the end of this year) and each experiment gathers the projected $\sim 30 \, {\rm fb^{-1}}$ before shutdown, we can study the NSUSY prospects for fits of this form by the end of the $8 \,{\rm TeV}$ run. We reduce each of the reported $8 \, {\rm TeV}$ signal strength errors by a factor 2, down to $\sigma_i/2$, and assume two different scenarios for the
final 8 TeV $\hat{\mu}_i$ central values: {\it 1)} they retain their current central values, or alternatively, {\it 2)} they move to $\hat{\mu}_i =1$. As the integrated luminosity is to increase by more than a factor of four while more channels are also expected to be added to the global
Higgs dataset, this estimate is conservative. We show the prospective best-fit regions in the two dimensional Wilson coefficient space of the $hgg$ and $h\gamma\gamma$ operators in these two hypothetical cases in Fig.~\ref{prospects}.
It is clear that a much more constrained stop space due to these Higgs fits, and possibly a $1-2 \sigma$ exclusion of NSUSY, will be feasible by the end of the $8 \,{\rm TeV}$ run.
Due to decoupling, if the SM emerges as a better fit in the dataset, with $\hat{\mu}_i^{8 \, {\rm TeV}} \rightarrow 1$, NSUSY will be relatively less constrained by fits of this form, however, fine-tuning arguments 
could be used against it. 

\subsection{Prospects for bounding the best-fit regions: collider and Higgs search exclusions} 

In this Section we compare the impact of the global fit on the stop parameter space with the searches based on missing energy.
The global fit we have performed with NSUSY draws interesting conclusions: the $\chi^2$ distribution is quite flat and nearly degenerate stops of mass as light as 400 GeV can offer a good fit to data (the degeneracy condition can even be relaxed in the absence of mixing). 
On the other hand, if the value of the Higgs mass is due solely to the stop sector, the allowed region shifts. Somewhat heavier stops, with a larger separation ($\delta m \gtrsim 500$ GeV) seem to be indicated in this case, mostly outside of the exceptional "funnel" region of parameter space. The same fit to Higgs data does rule out a small portion of the stop parameter space, comparable to the monophoton exclusion. Note however, that those two exclusions, from monophoton searches and from Higgs data, have different degrees of SUSY model dependence. The monophoton exclusion is based on missing energy signatures due to assumed R-parity conservation, whereas the Higgs data exclusion does not assume it,  nor any particular stop decay chain.

In the experimental searches, stops are assumed to decay to $\chi^{\pm,0}$. In the case of NSUSY, those electroweak states are degenerate higgsinos, leading in both cases to missing energy in association with a b-jet or a top,
\bea
\tilde{t} & \to & b \, \tilde{\chi}^{+} \to b \tilde{\chi}^0 + \textrm{ soft objects} \nonumber \\
\tilde{t} & \to & t \, \tilde{\chi}^0 \ .
\eea
The first decay chain corresponds to the direct sbottom search~\cite{direct-sbottom,Li:2010zv,Datta:2011ef,Lee:2012sy,Alvarez:2012wf}. The second decay contains a top (on- or off-shell), considered by the direct stop searches by ATLAS (See Section~\ref{collider-bounds}).

\begin{figure}[t]
\centering
\includegraphics[width=70mm,height=60mm]{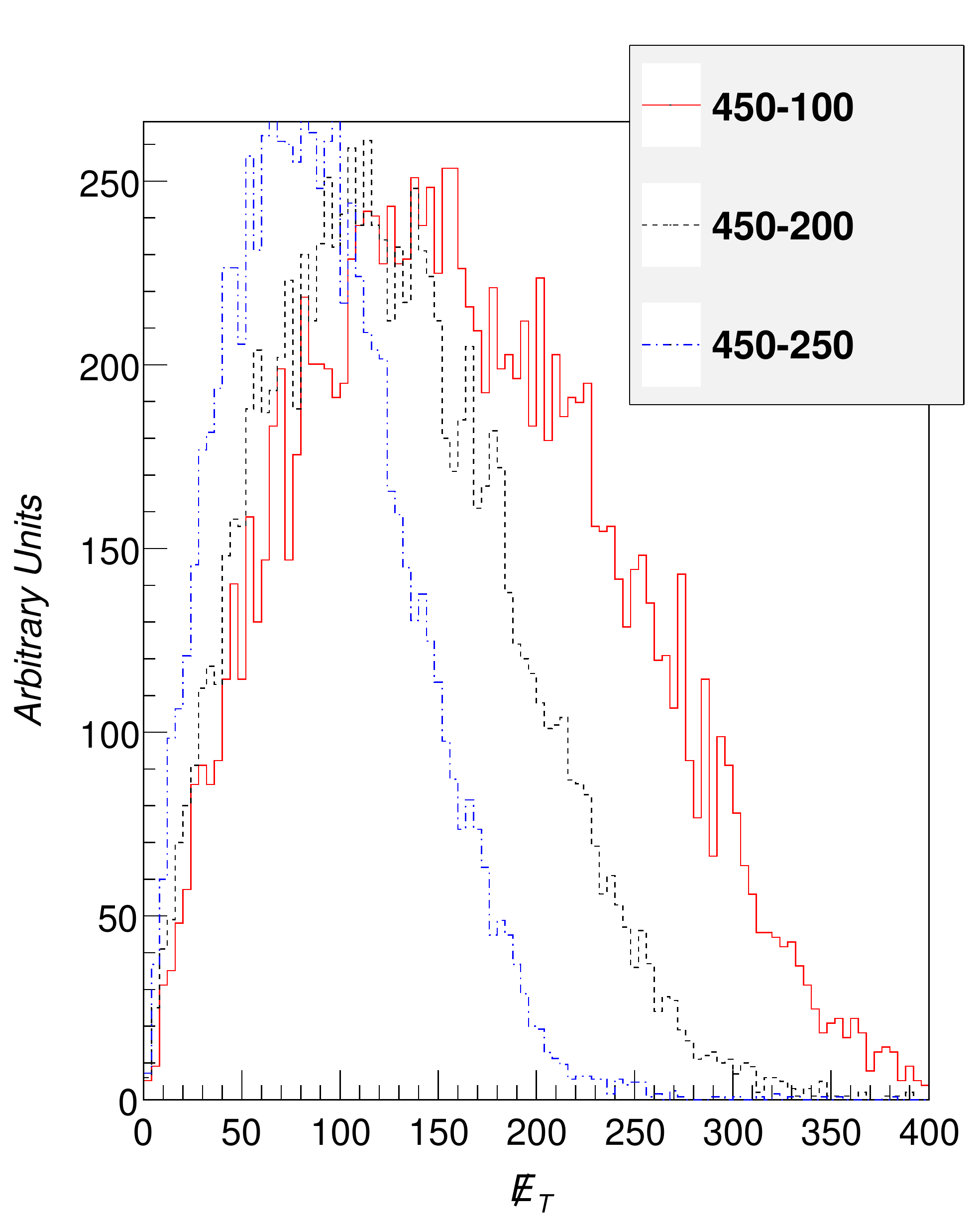}
\includegraphics[width=70mm,height=60mm]{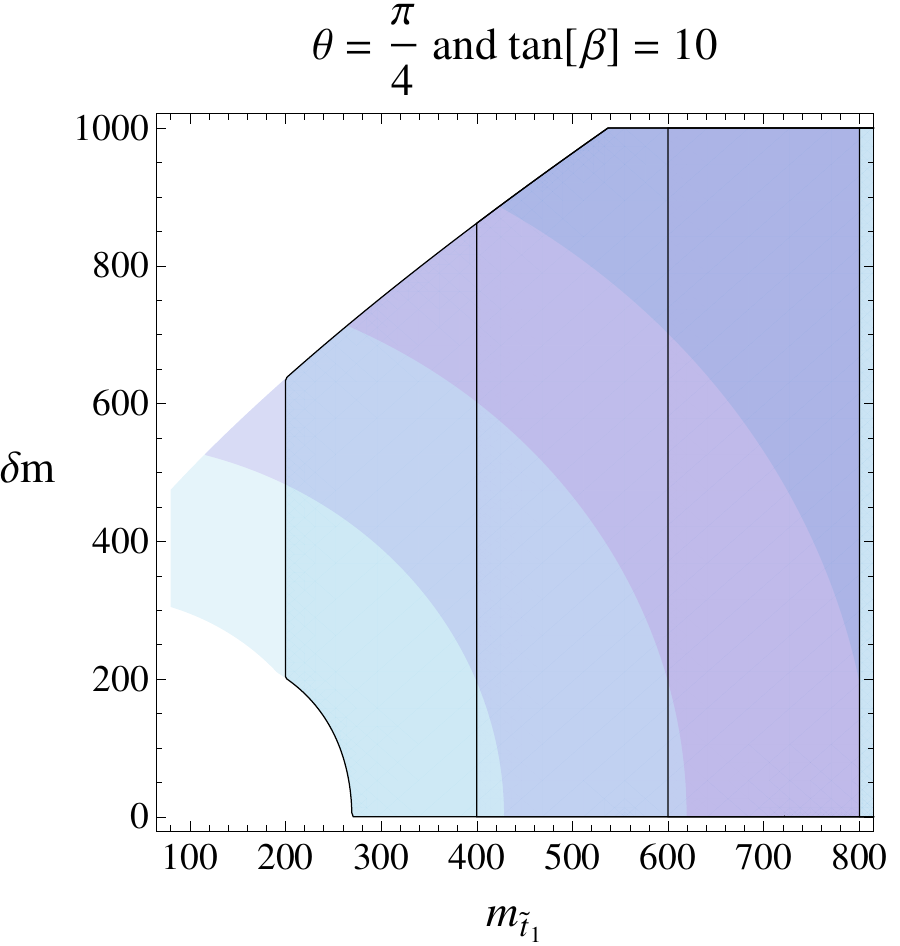}
\caption{Left figure:  The missing energy distribution for different choices of neutralino mass, with the stop mass fixed to 450 GeV. The red-solid, black-dotted and blue-dashed lines correspond to $m_{\tilde{\chi}}$=100, 200 and 250 GeV.  Notice the decrease of missing energy as the neutralino mass increases.  Right figure: Restriction in the stop space due the limits on the mass splitting of $\tilde{b}_L,\tilde{t}_L$, with a lower bound on $m_{\tilde{b}}>$ 200, 400, 600 and 800 GeV. The vertical lines correspond to interpreting direct sbottom searches as a $\tilde{t} \to b \tilde{\chi}^{\pm}$ decay, leading to the same mass limit.}
\label{MET}
\end{figure}

 The current stop searches are very weak in the region of $m_{\tilde{\chi}^0}\sim \mu > 90$ GeV. The current ATLAS exclusion region is a small triangle below $\mu=$160 GeV, and between, roughly, 300 and 450 GeV. To explore NSUSY in this best-fit region, one would need to push the stop searches not towards larger values of $m_{\tilde{t}}$, but of $m_{\tilde{\chi}^0}$. The main issue to reach higher values of $\mu$ is the similarity with $t\bar{t}$ either for $m_{\tilde{t}}\sim m_t+\mu$, or large $\mu$. In both cases, the missing energy distribution loses its ability to suppress the top background. The stop has little phase space to produce boosted $\chi^{\pm,0}$ unless it itself comes with some boost. For $m_{\tilde{t}}\sim$ 300 GeV, this would need some requirement on radiation jets. We illustrate the reduction of missing energy as one increases the value of $\mu$ in Fig.~\ref{MET}. Similarly, direct detection of charginos is difficult, due to  the degeneracy with the neutralino and small cross sections.  
 
Direct sbottom production searches can be translated into stop production searches when the stop decays to $b$ and $\chi^{\pm}$, and this search is more encouraging. The ATLAS collaboration performed a search for sbottom squarks, resulting in a 95\% C.L. upper limit $m_{\tilde b_1}>$ 390 GeV for neutralino masses below 60 GeV with ${\cal L}=2.05 {\rm fb}^{-1}$~\cite{Aad:2011cw}, but would probably be extended to larger values of $m_{\chi^0}$ with the 2012 dataset, being able to cover the case $m_{\tilde{\chi}^{0,\pm}}\lesssim$ 200 GeV. 
Note that direct sbottom searches assume ${\rm BR}(\tilde{b}\to b \tilde{\chi}^0)$=1 but can be re-scaled for lower branching ratios. Moreover, pushing the sbottom limits, within NSUSY, is also an indirect probe of the stop sector due to the custodial relations shown in Section~\ref{EWPD}. This interplay between direct and indirect limits on stops from direct sbottom production searches is illustrated in Fig.~\ref{MET}, where the maximal mixing case is shown. Interestingly, if the lightest stop is  L-dominated, the combination of sbottom and custodial violation limits could be more constraining than the re-interpretation of direct searches in terms of stop decays into charginos. As we do not know the chiral admixture of the stops and sbottoms, using both strategies would give the best sensitivity. Since sbottom direct detection searches do not suffer many of the problems that affect the advance in collider stop searches, this further supports the observation that direct sbottom searches may be the best direct access to stop
parameter space in the near term~\cite{Lee:2012sy}.
\begin{figure}[h!]
\centering
\includegraphics[scale=0.3]{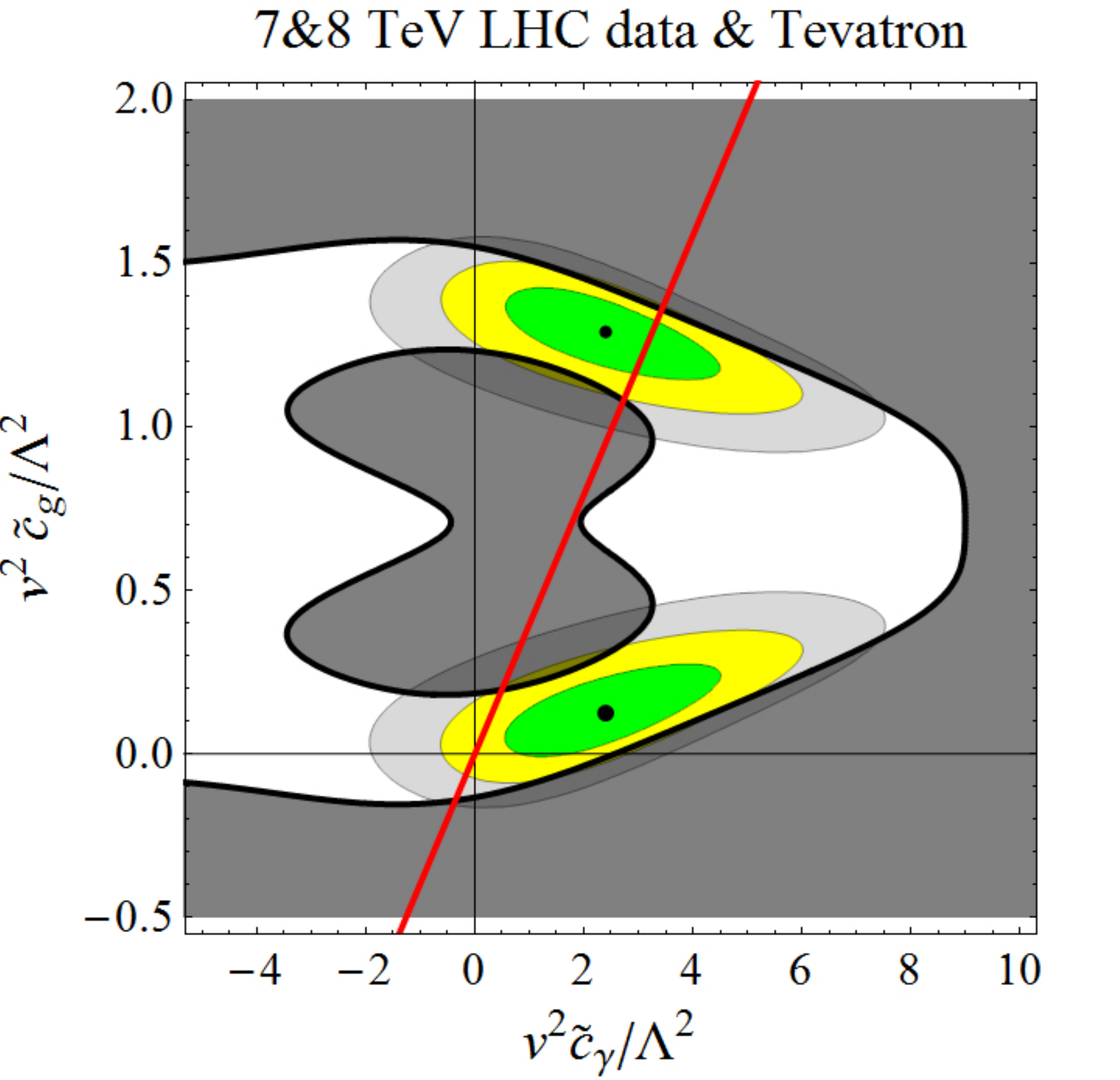}
\caption{The region of the higher dimensional operator space excluded at $95 \%$ C.L. from current Higgs searches is shown by the black shaded area overlaid on the best-fit regions. The NSUSY relation between the Wilson coefficients is shown with a red line.} 
\label{current-exclusion}
\end{figure}

\begin{figure}[h!]
\centering
\includegraphics[scale=1]{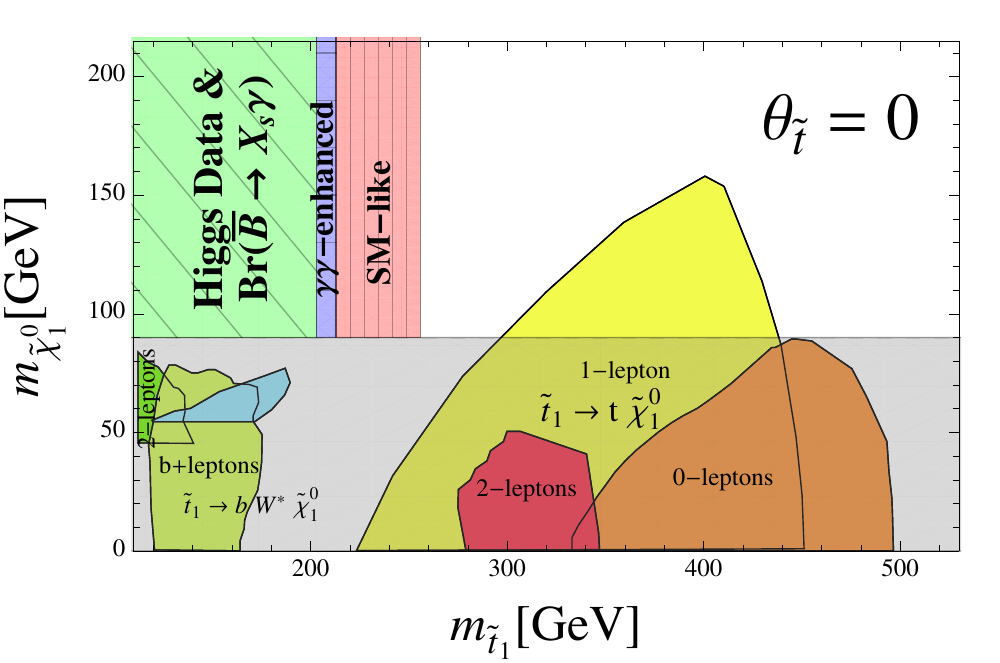}
\includegraphics[scale=1]{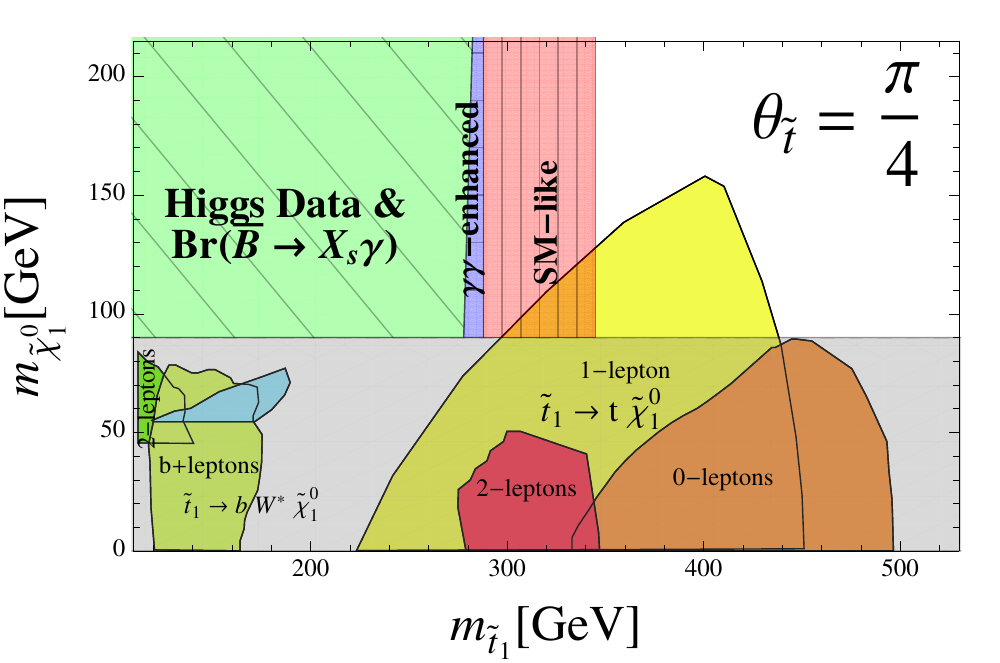}
\caption{The current (future) exclusion regions from current (and prospective) Higgs search data directly combined with the condition that  the 
${\rm Br}(\bar{B} \rightarrow X_s \, \gamma)$ constraint is within its $2 \, \sigma$ allowed region. Shown are the summary $95 \%$ C.L. exclusion results from the dedicated ATLAS collider searches for direct stop production in the $m_{\tilde{t}_1},\tilde{\chi}_1^0$ plane as solid regions (generally below the line $\tilde{\chi}_1^0 = 90 \, {\rm GeV}$). These results are reported in Ref.~\cite{direct-stop} and were combined into the presented form at ICHEP2012. Added to these figures in the $\tilde{\chi}_1^0 \simeq 90-200 \, {\rm GeV}$  space are current exclusions based on Higgs data and the ${\rm Br}(\bar{B} \rightarrow X_s \, \gamma)$ constraint  as a green region with diagonal lines; future exclusions from Higgs search data  that evolves to more SM-like values (red region with vertical lines, labelled "SM-like") or when the global Higgs data has evolved to have the current $8 \, {\rm TeV}$ central values while its errors have been reduced by a factor of $2$ (blue region, labelled "$\gamma\gamma$ enhanced"). Top (bottom) figure is the no (maximal) mixing case.}
\label{exclusion-prospect}
\end{figure}

Higgs search data also provides a powerful insight into stop parameter space. In particular they offer experimental reach into the large $\mu$ region that is such a challenge for collider searches. 
In Fig.~\ref{current-exclusion}, we show the region of the higher dimensional operator space excluded at $95 \%$ C.L. from 
current Higgs searches.\footnote{See Ref.~\cite{Espinosa:2012im} for more details on the exclusion methodology.} Translating this exclusion into the stop parameter space does not lead in general to
direct lower mass bound that is independent of $ \delta m$ (when $\theta_{\tilde{t}} \neq 0$)
due to the existence of the ``funnel region" where a cancelation of the stop contributions to $F_g$ can occur.
However, by jointly imposing the condition that the ${\rm Br}(\bar{B} \rightarrow X_s \, \gamma)$ constraint is within its $2 \, \sigma $ allowed region, we can define more stringent current exclusion regions for approximately mass degenerate stops.
By studying how the Higgs data will scale with more luminosity, we can also project expected exclusions in the stop space by the end of the $8 \, {\rm TeV}$ LHC run. 
We find the results shown in Fig~\ref{exclusion-prospect}. 


Note that these exclusions are in the region of $\mu \sim 100-200 \, {\rm GeV}$ in NSUSY and do not require a missing energy tag. This makes indirect studies of stop exclusions from Higgs search data broadly applicable to many SUSY models and these bounds have significant reach into the large $\mu$ region. Conversely, collider based searches will 
be severely challenged to reach the large $\mu$ space in direct stop searches due to the fact that the signal region is increasingly
similar to the $t \, \bar{t}$ SM backgrounds with small missing energy.

\section{Conclusions.}\label{concl}

In this paper we have examined the light that Higgs search data sheds on the stop sector of Natural SUSY, where
stops are expected to lead to the largest effects on Higgs properties. We have 
interpreted a generic fit to the latest Higgs data in models which allow for 
additional BSM contributions to the $hgg$ and $h\gamma\gamma$ couplings in terms of stop parameters.
In performing this careful study of the impact of light stop states on Higgs properties, we have included QCD matching corrections and  consistently treated the allowed best-fit regions with a one parameter CDF.  Further, we have examined in detail the impact of related precision flavour and EW constraints. 

Interestingly, we find that a combined global fit to Higgs properties, ${\rm Br}(\bar{B} \rightarrow X_s \, \gamma)$ and $\Delta m_W$,  show a mild preference for $\sim 400-500 \, {\rm GeV}$ degenerate stops in the no mixing case when $\mu \sim 100 \, {\rm GeV}$. Such a spectrum of stop states cannot explain directly the Higgs mass value of $\sim 125 \, {\rm GeV}$ but does offer a good fit to the global dataset we have studied -- the quality of fit is comparable to the goodness of fit in the SM. Such a stop spectrum can also improve the agreement of the observed and predicted $m_W$, ameliorating a very slight tension in the SM with this observable that has arisen with recent precise measurements at the Tevatron. However, we emphasize that what is more interesting at this time is the degree of consistency that the small shift in this precision observable has with the allowed low mass region in the Higgs global fits.
Future careful studies of this form will be essential in ruling out, or confirming a discovery of Natural SUSY at the LHC.

\appendix*

\section{}

\subsection{SM Inputs Used}
The SM inputs used are shown in Table \ref{table:table2}.

\begin{table}[h] 
\setlength{\tabcolsep}{5pt}
\center
\begin{tabular}{c|c} 
\hline \hline 
Input & Value \\
\hline
$m_h$ & $125 \pm 2 \, {\rm GeV}$ \\
$m_t$ & $173.2 \pm 0.9 \, {\rm GeV}$ \cite{Lancaster:2011wr} \\
$m_b({\rm 1S})$ & $4.65 \pm 0.03 \, {\rm GeV}$ \cite{Bauer:2004ve,Nakamura:2010zzi} \\
$m_c$ & $1.275 \pm 0.025 \,  {\rm GeV} $ \cite{Nakamura:2010zzi} \\
$m_\tau$ & $1776.82 � 0.16 \, {\rm MeV}$ \cite{Nakamura:2010zzi} \\
$m_Z$ & $91.1876 \pm 0.0021\, {\rm GeV}$ \cite{Nakamura:2010zzi}  \\
$\alpha_s(M_Z)$ & $0.1184 \pm 0.0007 \, {\rm GeV}$ \cite{Bethke:2009jm}\\
$(\Delta \alpha)^{(5)}_{had}$ &  $\left(275.7 \pm 1\right) \times 10^{-4}$ \cite{Davier:2010nc} \\
$(\Delta \alpha)_{lep}$ &  $\left(314.97686 \right) \times 10^{-4}$  \cite{Steinhauser:1998rq} \\  
$s_W^2$ &  $0.2233 \pm 0.0002$ \cite{Nakamura:2010zzi} \\
$G_F$ & $1.166 378 7 \pm 0.0000006  \times 10^{-5} \, {\rm GeV^{-2}}$\cite{Nakamura:2010zzi}  \\
 \hline \hline
\end{tabular}
\caption{\it Input values used in determining constraints on the NSUSY parameters in the indirect tests. When we use the value $\alpha_s(m_t)$ we determine the value from NLO running
in QCD finding the central value $\alpha_s(m_t) = 0.1080$.}
\label{table:table2} \vspace{-0.35cm}
\end{table}

\subsection{Efficiency corrections due to Higher Dimensional Operators}
We seek to  draw as precise a conclusion as possible about the Higgs signal strength fit to higher dimensional operators with Wilson coefficients $\tilde{c}_g,\tilde{c}_{\gamma}$ in this paper.
The effect of these higher dimensional operators on the signal strength parameters can have two forms. Directly the $\sigma_{gg \rightarrow h}$ and $\Gamma_{h \rightarrow \gamma \, \gamma}$ rates can be effected,
as characterized by our rescaling the data with the effects of $\tilde{c}_g,\tilde{c}_{\gamma}$ consistently. This effect is what we fit to using our global fit procedure. 

There is also a further effect that higher dimensional operators can have on the event yields that lead to the $\mu_i$. The higher dimensional operators can alter the shape of the final
kinematic distributions (that sum over more than one production mechanism), leading to a further correction, as a function of the Wilson coefficient, on the signal strength parameter. The $\tilde{c}_g,\tilde{c}_{\gamma}$ do not affect the shape of the kinematic distributions due to of an individual production process, such as $gg \rightarrow h \rightarrow \gamma \, \gamma$, however, they do affect the relative proportions of $gg$ versus Higgstrahlung and VBF initiated Higgs production.
This later effect is an ``efficiency correction" on the change in the number of expected events expected due to a different effective efficiency for passing the cuts of the experimental analyses 
when higher dimensional operators are present. This effect is not captured in our direct fit to $\tilde{c}_g,\tilde{c}_{\gamma}$ and should be quantified for precise conclusions.

We have examined the effect of these efficiency corrections through simulating the effect of those operators in the 7 TeV run data in the $p \, p \to h \to \gamma\gamma \, (0j, 1j, 2 j)$. We added the shift due to the operators in the $h g g$  and $h \gamma \gamma$ effective vertices using Feynrules~\cite{feynrules} into an UFO model~\cite{UFO} of MadGraph5~\cite{Alwall:2011uj}. We generated samples at 7 TeV with a parton level cut on $|\eta_{\gamma}|<$ 3.5 and $p_{T,\gamma}>$ 20 GeV.  
The parton level events generated by MadGraph5 are passed to Pythia~\cite{Sjostrand:2006za} to simulate the effects of parton showering, and then to Delphes~\cite{Ovyn:2009tx} for a fast detector simulation. We use the generic LHC parameters for Delphes, and reconstruct jets with the anti-$k_T$ algorithm using $0.5$ for jet cone radius. We cannot simulate all the characteristics of photons in the different bins, especially the converted/unconverted nature of the photon. Instead, we check the effect of the new physics on the  basic selection cuts described in Table~\ref{cuts}. Those are on top of the detector effects simulated by Delphes, including isolation and energy-momentum smearing.
\begin{table}[h] 
\setlength{\tabcolsep}{5pt}
\center
\begin{tabular}{c|c} 
\hline \hline 
ATLAS cuts & CMS cuts \\ 
\hline
$|\eta_{\gamma_{1,2}}|< 2.37$ &   $|\eta_{\gamma_{1,2}}|< 2.5 $ \\ 
$p_{T}^{\gamma_1} > 40$ GeV,  $p_{T}^{\gamma_2} > 25$ (30) GeV & $p_{T}^{\gamma_1} > m_{\gamma\gamma}/3$,  $p_{T}^{\gamma_2} > m_{\gamma\gamma}/4$ \\
$m_{\gamma\gamma} = 125 (126) \pm 3$ GeV  &  $m_{\gamma\gamma} = 125 \pm 3$ GeV  \\ \hline\hline
\end{tabular}
\caption{\it Cuts implemented in the Monte Carlo simulation after Delphes. Numbers in parenthesis correspond to the 8 TeV run.}
\label{cuts}
 \vspace{-0.35cm}
\end{table}

\begin{figure}[h!]
\centering
\includegraphics[scale=0.8]{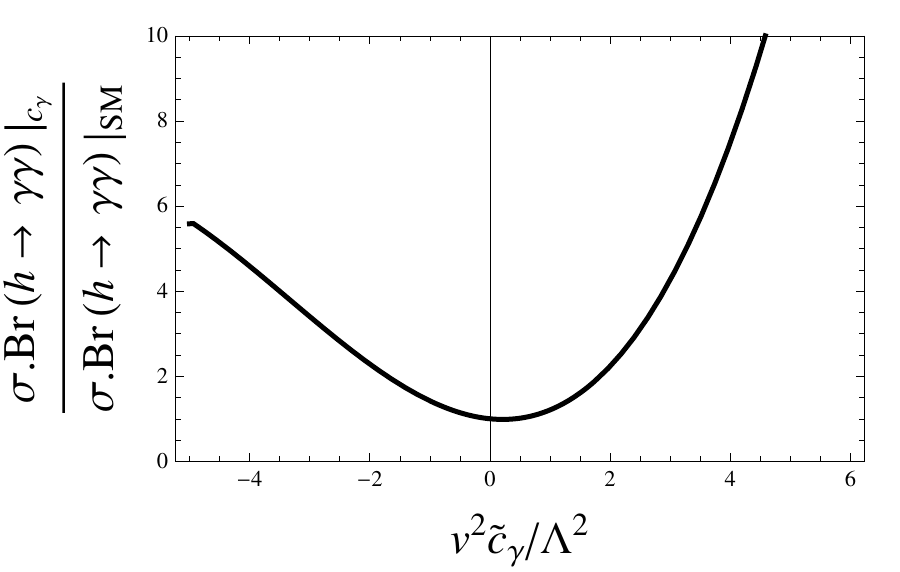}
\includegraphics[scale=0.16]{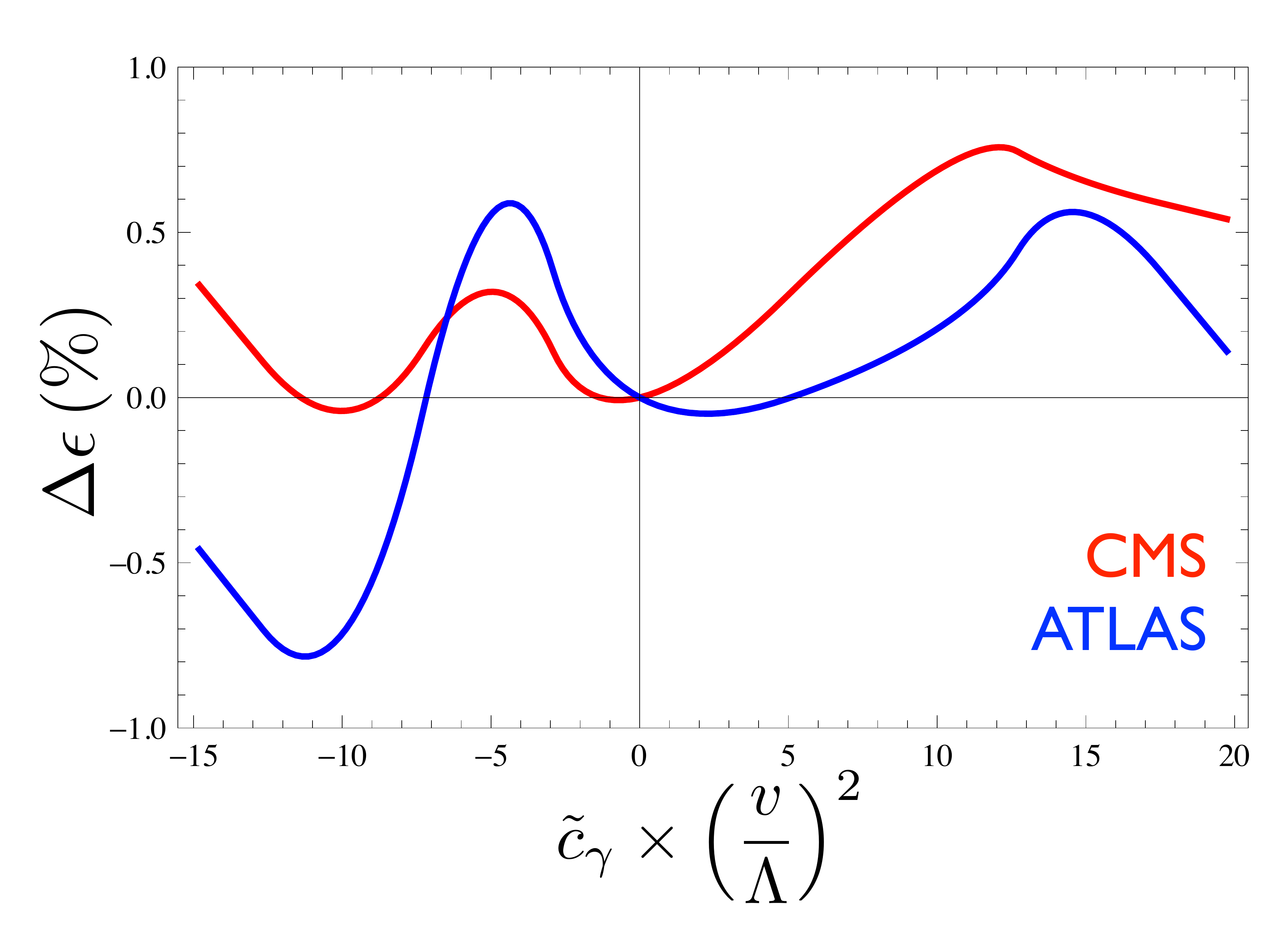}
\caption{Left: The cross section as a function of $c_{\gamma}$, $c_{g}$ using the fixed relationship between the Wilson coefficients. Right: Change in efficiencies to the cuts defined in Table.~\ref{cuts}. Both figures correspond to $\sqrt{s}=$ 7 TeV.}
\label{effcgam}
\end{figure}

Those cuts are then applied to the signal, containing the SM and the new operators. In Fig.~\ref{effcgam} we show the effect of  the operators, $c_{\gamma}, c_g$. The effect on the cross section is sizeable, whereas the effect on the efficiencies due to basic $p_T$, $\eta$ and $\Delta m_{\gamma\gamma}$ cuts is moderate (less that 1\%) except when the cross section drops due to a cancellation between the SM and new physics contributions.
 
The effect on efficiencies for the relation between $c_{\gamma}$ and $c_{g}$ given in Eq.~\ref{rel-stop} is very similar to Fig.~\ref{effcgam}, namely of the order $\lesssim$ 1 \%.
Further the corrections stay at this level when both operators are present or when the effects at $8 \, {\rm TeV}$ are simulated as we have explicitly verified.
Due to this small correction we neglect effects due to efficiency corrections due to higher dimensional operators in this paper. 
\subsection{${\bma{{\rm Br}(\bar{B} \rightarrow X_s \, \gamma)}}$ Loop Functions}
\label{bsg-func}
The loop functions in ${\rm Br}(\bar{B} \rightarrow X_s \, \gamma)$ are given by~\cite{Bertolini:1990if,Grzadkowski:2008mf,Degrassi:2000qf}
\bea
F_{7}^1(x) &=& \frac{x \, (7 - 5 x - 8 x^2)}{24\,(x-1)^3} + \frac{x^2 \, ( 3 x - 2)}{4 (x-1)^4} \, \log x,  \\
F_{8}^1(x) &=& \frac{x \, (2 + 5 x - x^2)}{8\,(x-1)^3} - \frac{3 x^2}{4 (x-1)^4} \, \log x, \\
F_{7}^3(x) &=& \frac{5 - 7 x}{6\,(x-1)^2} + \frac{x \, ( 3 x - 2)}{3 (x-1)^3} \, \log x,  \\
F_{8}^3(x) &=& \frac{1 + x}{2\,(x-1)^2} - \frac{x}{(x-1)^3} \, \log x.
\eea
Note the sign correction in $F_{8}^1(x)$ when comparing to the previous version of this paper.

\begin{figure}[h!]
\centering
\includegraphics[scale=0.67]{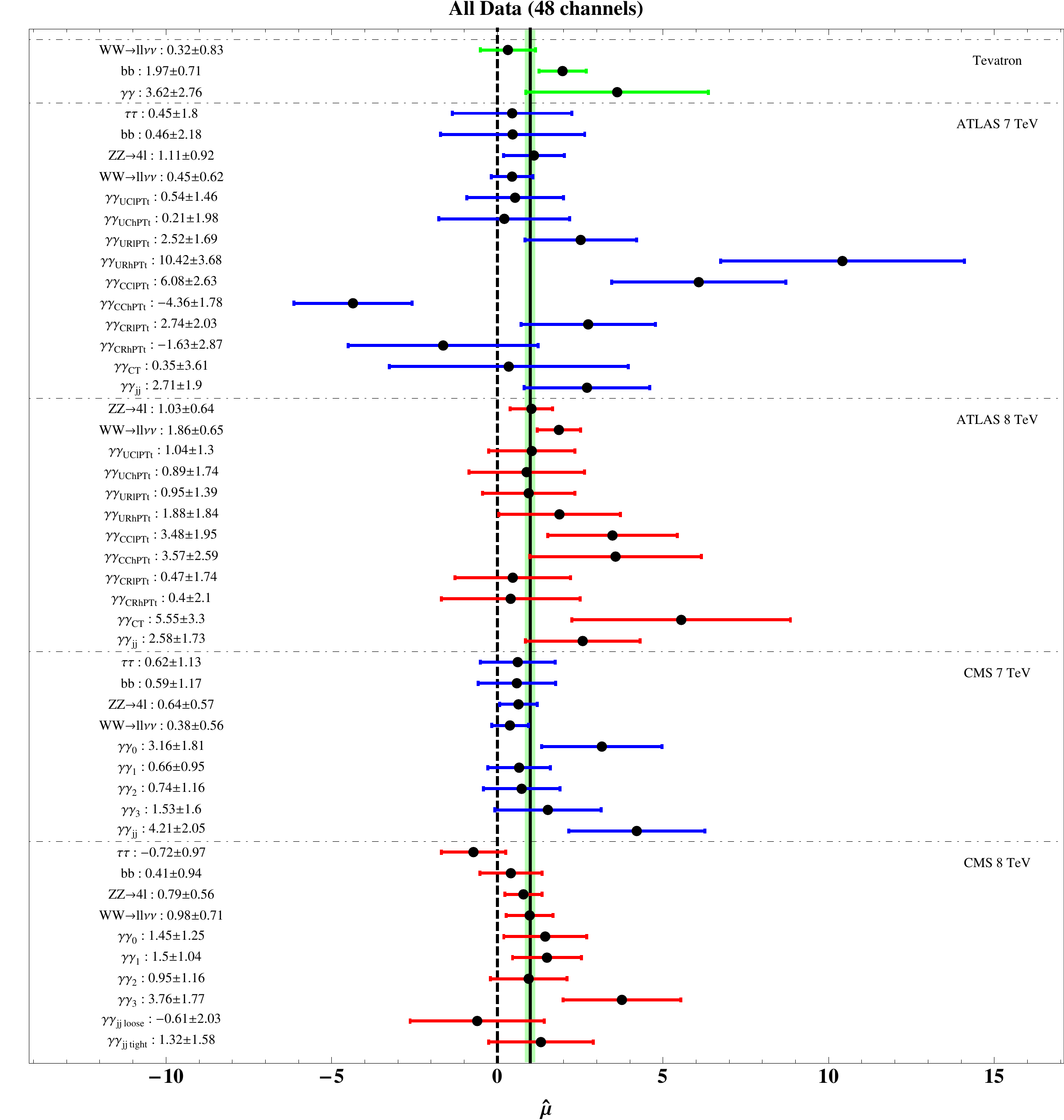}
\caption{Summary of Higgs best-fit signal-strengths $\hat\mu_i$ used in our global fit. For more details see
Ref.~\cite{Espinosa:2012ir,Espinosa:2012vu,Espinosa:2012im}. Shown are the reported $\hat\mu_i$ (or the reconstructed $8 \, {\rm TeV}$ value if not directly reported \cite{Espinosa:2012im}). For CMS and the Tevatron we use values at $m_h  = 125 \, {\rm GeV}$, while for ATLAS we use $m_h = 126.5 \, {\rm GeV}$. This choice is partially due to the limited experimental information currently supplied. 
 For discussions on possible bias from this choice see Ref.~\cite{Espinosa:2012im,Murray}.
Also shown is the combined $\hat\mu$ and error as a vertical green band and the SM expected signal strength as a black vertical line at $\hat{\mu} =1$.}
\label{higgsdata}
\end{figure}

\subsection*{Acknowledgments}
We particularly thank M. M\"uhlleitner for collaboration developing the global fit. We also thank K. ~Tackman, P.~Slavich, B.~Feigl, R.~Harlander, M.~Schumacher, M.~Spira, W.~Fisher, J.~Huston,
V.~Sharma, P.~Uwer, J.~Bendavid, J.~Virto and G. Isidori for helpful communication concerning both theory and data related to the global fit and analysis in this paper.
VS would like to thank Xavier Portell for explaining aspects of stop searches at ATLAS. 
This work has been partly supported by the European Commission under the contract ERC advanced
grant 226371 �MassTeV�, the contract PITN-GA-2009-237920 �UNILHC�, and
the contract MRTN-CT-2006-035863 �ForcesUniverse�, as well as by the
Spanish Consolider Ingenio 2010 Programme CPAN (CSD2007-00042) and the
Spanish Ministry MICNN under contract FPA2010-17747 and
FPA2011-25948.


\end{document}